\documentclass[longtitle,preprint,5p,times,twocolumn]{elsarticle}

\usepackage{amssymb}

\usepackage{caption}
\usepackage{xspace}
\usepackage{longtable}
\usepackage{pdflscape}
\usepackage{multirow}
\usepackage{ragged2e}
\usepackage{booktabs}
\usepackage{hyperref}
\usepackage{lineno}

\def\notenumber{ecce-paper-phys-2022-08}
\def\noteversion{v1.0}

\begin{document}
\begin{frontmatter}

\title{ECCE Sensitivity Studies for Single Hadron\\Transverse Single Spin Asymmetry Measurements}
\tnotetext[t1]{
	    \includegraphics[scale=0.075]{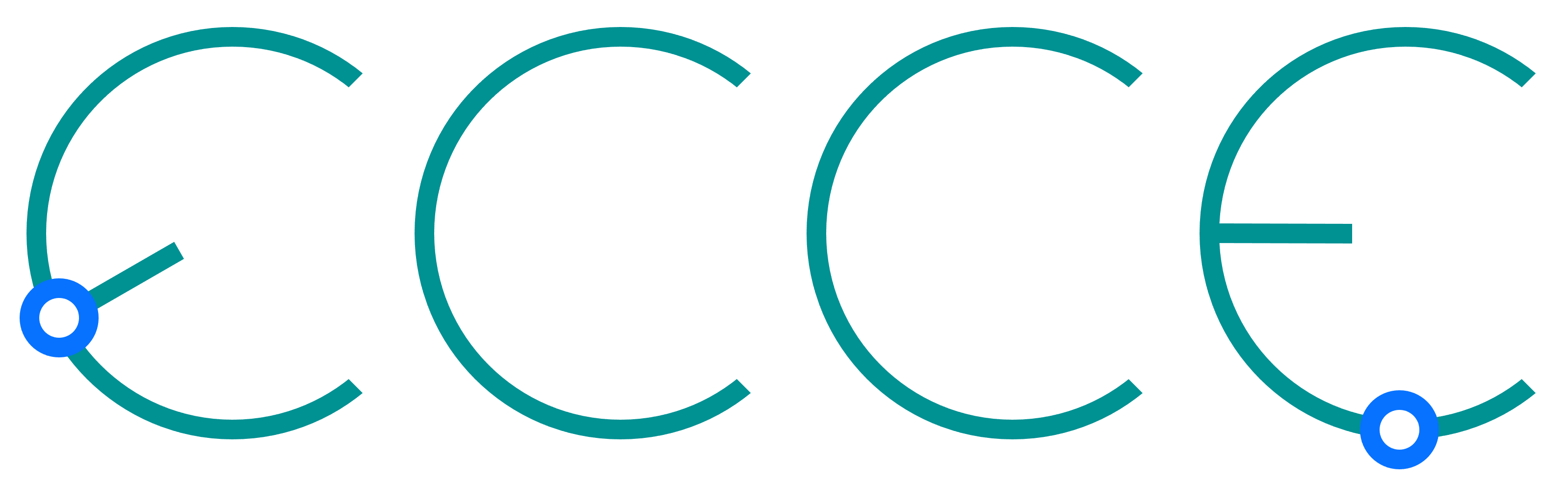} \\
	    \notenumber \\
	    \noteversion \\
	    \today	\\
}
\def\theaffn{\arabic{affn}} 
\author[RIKEN,RBRC]{Ralf Seidl}
\author[Regensburg,Madrid]{Alexey Vladimirov}
\author[Lebanon]{Daniel Pitonyak}
\author[PSB]{Alexei Prokudin}

%
%
%
%

\author[MoreheadState]{J.~K.~Adkins}
\author[RIKEN,RBRC]{Y.~Akiba}
\author[UKansas]{A.~Albataineh}
\author[ODU]{M.~Amaryan}
\author[Oslo]{I.~C.~Arsene}
\author[MSU]{C. Ayerbe Gayoso}
% ...

%

% ...
\affiliation[MoreheadState]{organization={Morehead State University},
	 city={Morehead},
	 state={KY},
	 }
 
% ...
\affiliation[MSU]{organization={Mississippi State University},
	 city={Mississippi State},
	 state={MS},
	 country={USA}} 
 
% ...
\affiliation[ODU]{organization={Old Dominion University},
	 city={Norfolk},
	 state={VA},
	 country={USA}} 
 
% ...
\affiliation[RIKEN]{organization={RIKEN Nishina Center},
	 city={Wako},
	 state={Saitama},
	 country={Japan}} 
 
% ...
\affiliation[RBRC]{organization={RIKEN BNL Research Center},
	 city={Upton},
	 state={NY},
	 country={USA}} 
	 
% ...
\affiliation[UKansas]{organization={Unviersity of Kansas},
	 city={Lawrence},
	 state={KS},
	 country={USA}} 
 
% ...
\affiliation[Oslo]{organization={University of Oslo},
	 city={Oslo},
	 country={Norway}} 
 
% ...
\affiliation[Lebanon]{organization={Lebanon Valley College},
	 city={Annville},
	 postcode={},
	 state={PA},
	 country={USA}}

\affiliation[PSB]{organization={Penn State University Berks},
	 city={Reading},
	 postcode={},
	 state={PA},
	 country={USA}} 
 
\affiliation[Regensburg]{organization={Universit"at Regensburg},
	 city={Regensburg},
	 	 postcode={},
	 country={Germany}} 
	 
	 \affiliation[Madrid]{organization={Universidad Complutense de Madrid},
city={Madrid},
postcode={ E-28040},
country={Spain}}

    \begin{abstract}
        We performed feasibility studies for various single transverse spin measurements that are related to the Sivers effect, transversity and the tensor charge, and the Collins fragmentation function. The processes studied include semi-inclusive deep inelastic scattering (SIDIS) where single hadrons (pions and kaons) were detected in addition to the scattered DIS lepton. The data were obtained in {\sc pythia}6 and {\sc geant}4 simulated e+p collisions at 18 GeV on 275 GeV, 18 on 100, 10 on 100, and 5 on 41 that use the ECCE detector configuration. Typical DIS kinematics were selected, most notably $Q^2 > 1 $ GeV$^2$, and cover the $x$ range from $10^{-4}$ to $1$. The single spin asymmetries were extracted as a function of $x$ and $Q^2$, as well as the semi-inclusive variables $z$, which corresponds to the momentum fraction the detected hadron carries relative to the struck parton, and $P_T$, which corresponds to the transverse momentum of the detected hadron relative to the virtual photon. They are obtained in azimuthal moments in combinations of the azimuthal angles of the hadron transverse momentum and transverse spin of the nucleon relative to the lepton scattering plane. In order to extract asymmetries, the initially unpolarized MonteCarlo was re-weighted in the true kinematic variables, hadron types and parton flavors based on global fits of fixed target SIDIS experiments and $e^+e^-$ annihilation data. The expected statistical precision of such measurements is extrapolated to 10 fb$^{-1}$ and potential systematic uncertainties are approximated given the deviations between true and reconstructed yields. Similar neutron information is obtained by comparing the ECCE e+p pseudo-data with the same from the EIC Yellow Report and scaling the corresponding Yellow Report e+$^3$He pseudo-data uncertainties accordingly. The impact on the knowledge of the Sivers functions, transversity and tensor charges, and the Collins function has then been evaluated in the same phenomenological extractions as in the Yellow Report. The impact is found to be comparable to that obtained with the parameterized Yellow Report detector and shows that the ECCE detector configuration can fulfill the physics goals on these quantities. 

    \end{abstract}
\date{\today}

\end{frontmatter}





\section{Introduction}
\label{s:intro}
Historically, transverse single spin asymmetries have been the key to access the transverse momentum structure of the nucleon. Both of the most famous effects, the Sivers \cite{Sivers:1989cc} effect and the Collins \cite{Collins:1992kk} effect were initially suggested to describe the nonzero single spin asymmetries that were observed for pions by the E704 experiment in fixed-target proton-proton collisions \cite{FNAL-E704:1991ovg}.  While in the end not directly applicable to those processes, single spin asymmetries of semi-inclusive processes did provide indications of nonzero asymmetries. In addition, both of these effects were clearly identified by the HERMES experiment in 2004\cite{Airapetian:2004tw}. Since then these results have been confirmed in more detail by HERMES \cite{HERMES:2013quo}, COMPASS \cite{COMPASS:2008isr,COMPASS:2014bze}, JLAB \cite{JeffersonLabHallA:2011ayy}, as well as various $e^+e^-$ results \cite{Seidl:2005zx,Belle:2008fdv,TheBABAR:2013yha,BESIII:2015fyw}. 


\begin{figure*}
    \centering
    \includegraphics[width=0.9\textwidth]{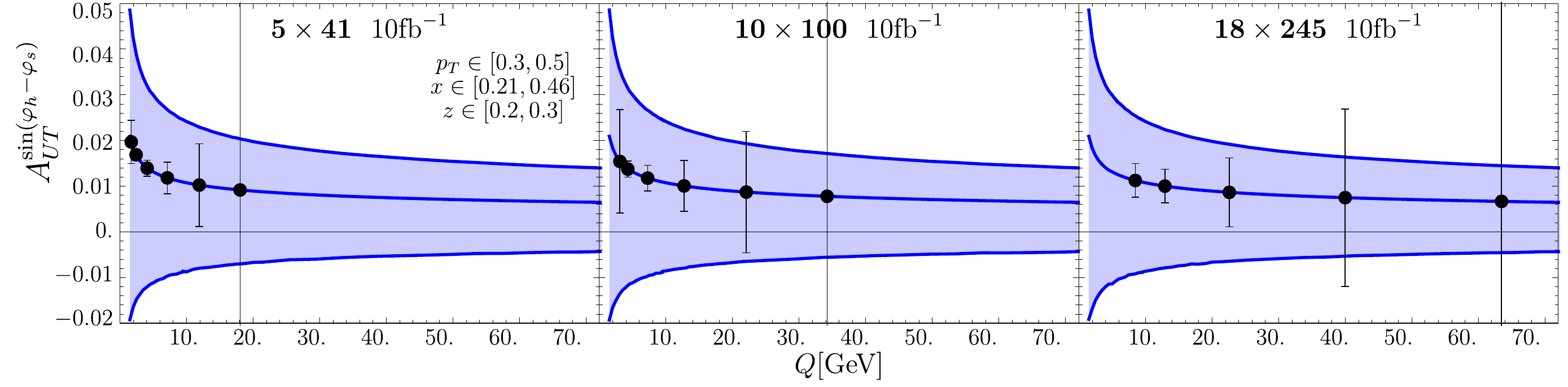}
    \caption{Example of the expected evolution effects from \cite{Bury:2021sue} for the Sivers asymmetry at an intermediate $x$, $z$ and $P_T$ value, as a function of $Q$ for three collision energy combinations. The error bands represent the current level of uncertainties and the data points represent the projected ECCE uncertainties (statistical and systematic uncertainties combined) to be discussed further below put to the central values of the current parameterization.}
    \label{fig:Sivevo}
\end{figure*}

These two effects are related to the corresponding transverse momentum depdendent distribution and fragmentation functions (TMDs) which give access to the three-dimensional momentum structure of the nucleon and provide some of the main information to spin-orbit effects of the nucleon. The Sivers function describes the correlation between the transverse spin of the nucleon and the transverse momentum of a parton within it. With the expected statistics and energy range of the EIC, the precise transverse and longitudinal momentum dependent distributions of not only valence but also sea quarks and gluons can be extracted. At present only up and down quark Sivers functions are known in the valence region but with rather large uncertainties
, particularly in the transverse momentum. Closely related to it is the scale dependence of the corresponding single spin asymmetries that is at present poorly known due to the fact that only fixed-target experimental data is available with very similar hard scales. Unlike the collinear case where at not too small momentum fractions $x$, the well-known DGLAP \cite{Dokshitzer:1977sg,Gribov:1972ri,Altarelli:1977zs} evolution is applicable, TMD evolution, especially at low scales again relies on universal functions that encode details of non-perturbative QCD dynamics and, at present, are mostly unknown. The future precise EIC data by both unpolarized and polarized TMD measurements will allow us to constrain the uncertainty in the TMD evolution. While it is a common misconception that the nonzero single spin asymmetries may disappear at higher scales, in all existing fits of TMD evolution effects, a logarithmic scale dependence is seen that could reduce the overall size of the asymmetries. 
In Fig.~\ref{fig:Sivevo} a kinematic example based on the \cite{Bury:2021sue} parameterization shows that low-scale Sivers asymmetries of about 2 \% would decrease to the sub-percent level at higher scales. As such, it is important for any EIC experiment to be able to reconstruct such asymmetries with both statistical and systematic precision below the 1 \% level over a large kinematic range in a fine enough binning. The details of the expected precision of the ECCE measurements will be discussed below, but one can already see the complementarity between different collision energies in covering a large lever arm with sufficient precision. 

The Collins effect \cite{Collins:1992kk} relates the chiral-odd quark transversity distribution \cite{Ralston:1979ys}, that is the basis for the tensor charge, with a polarized fragmentation function, the Collins fragmentation function. It correlates the transverse spin of a fragmenting parton with the azimuthal yield of final-state hadrons around the axis of this parton. Unlike the Sivers function, that can be accessed with an unpolarized fragmentation function, the fact that the fragmentation function is also polarized and chiral-odd makes the transversity extraction more difficult. Nevertheless, access to only the Collins FFs has been obtained from $e^+e^-$ annihilation measurements, initially by Belle \cite{Seidl:2005zx,Belle:2008fdv} and later by BABAR \cite{TheBABAR:2013yha} and BESIII\cite{BESIII:2015fyw}. 
Using this information together with the SIDIS data from HERMES, COMPASS and JLAB, various transversity extractions have been performed, although they predominantly rely on only valence flavors so far. Recently, also single-hadron single spin asymmetries from hadronic collisions were included in a global QCD analyssi of all avialable data on transverse spin asymmetries, including apart from SIDIS, Drell-Yan and $e^+e^-$ data also $A_N$ data from proton-proton scattering \cite{Cammarota:2020qcw}. 
The interest in the tensor charges stems from the fact that various interactions beyond the standard model may be also a tensor type of interaction \cite{Courtoy:2015haa}. As at the same time Lattice QCD calculations argue to be already fairly reliable on the calculation of the tensor charge, any discrepancies between measurement and Lattice results may indicate BSM effects. Although the tensor charges are expected to be more of a valence quark effect (due the the charges being defined as the difference of quark and antiquark transversities), fixed target measurements will not be able to perform the integral over large enough of an $x$ range to satisfactorily extract the charges, but the EIC can \cite{Gamberg:2021lgx}. Also here the scale dependence is of interest as well as accessing the sea quark transversity distributions.

\section{Data selection}

The simulated data were obtained using the pythiaeRHIC \cite{pythiaerhic} implementation of {\sc pythia}6 \cite{Sjostrand:2006za} with the same settings and events that were also used in the SIDIS studies of the EIC Yellow report \cite{EICYellowReport}. It should be noted that for these studies no dedicated radiative effects were generated other than what is already included in {\sc pythia}. These effects are likely very relevant, especially at large $y$ but are common to all EIC detector proposals and were therefore not studied here. 
The generated data, in its eic-smear \cite{eicsmear} format, was then run through a {\sc geant}4 simulation of ECCE that contains all the relevant tracking detectors and calorimeters, as well as some of the support material, magnet yoke, the PID detectors, etc., c.f. \cite{ecce-sim}. The PID information in these simulations came from a parametrization based on the rapidity and momentum dependent PID resolutions that can be expected for the various PID subsystems.   

The data was obtained at the energy combinations that are summarized in Table \ref{tab:lumi} where the simulations were separated into low $Q^2$ data and higher $Q^2$ data in order to still obtain reasonable statistics at the lower cross sections at higher $Q^2$. Unlike in the Yellow Report, no dedicated e+$^3$He
 simulations were run and instead for the impact studies the Yellow Report uncertainties were rescaled based on the ECCE e+p simulations. 
 As can be seen from these luminosities, especially at low $Q^2$ the accumulated data is still far below the level of statistics to be expected from the EIC. Nevertheless the statistics are large enough to evaluate the statistical uncertainties that can be expected. At the higher $Q^2>100$ GeV$^2$ range, the luminosities are generally larger which in turn compensates for the lower cross sections and event rates expected there. 
 
\begin{table}[htb]
    \centering
    \begin{tabular}{c | c  | r | c }
        Energy &  $Q^2$ range & events & Luminosity (fb$^{-1}$) \\\hline
          18x275 & 1 - 100 & 38.71M & 0.044 \\
               &  $>$  100 & 3.81M & 1.232 \\   
        18x100 & 1 - 100 & 14.92M & 0.022 \\
               &  $>$  100 & 3.72M & 2.147 \\
        10x100 & 1 - 100 & 39.02M & 0.067 \\
               &  $>$  100 & 1.89M & 1.631 \\
         5x41  & 1 - 100 & 39.18M & 0.123 \\
               &  $>$  100 & 0.96M & 5.944 \\\hline
    \end{tabular}
    \caption{MC statistics and luminosities used for the single spin asymmetry simulations. Part of the lower $Q^2$ range data was obtained from simulations without upper $Q^2$ cut.}
    \label{tab:lumi}
\end{table}

\section{General (SI)DIS kinematics, requirements}
As with all deeply inelastic scattering events the typical requirements on DIS kinematics are considered. The most important one is on the scale of the process by having a lower limit on the squared momentum transfer from the lepton to the nucleon, $Q^2 > 1$ GeV$^2$. Additionally, also the invariant mass of the hadronic final state is supposed to be above the main nucleon resonances which is ensured with $W^2 > 10$ GeV$^2$. 
Further requirements are made on the inelasticity to be $0.01 < y < 0.95$ where the lower criterion is motivated by the large smearing in the scattered lepton DIS kinematic reconstruction method and the upper limit is motivated by the large increase in radiative effects close to unity. 

For these studies the DIS kinematics are obtained by using the reconstructed scattered lepton kinematics which were solely obtained from the tracking. In future studies the addition of often better kinematic resolution from some of the elctromagnetic calorimeters may further improve the DIS and SIDIS resolutions. 
It was found that for most of the kinematic range the lepton method is reliable enough to obtain the relevant single spin asymmetries. Particularly at higher scales, the double angle and Jaquet-Blondel methods may help the resolutions and thus improve the systematic uncertainties even further. Additionally, using only the hadronic final state as in the JB method, one can make use also of charged-current reactions that provide even further flavor separation via the weak interaction. These topics are of additional interest and need to be studied in greater detail at a later time. 

In addition to the scattered lepton, at least one final-state hadron has to be detected in the main ECCE detector system ($|\eta|<3.5)$ that has been identified as either charged pion or kaon. While the particle identification within the ECCE detector will not be perfect over the whole range, for the single spin asymmetry studies we assumed that the PID efficiencies will be known well enough to be reliably unfolded.

For this analysis a multi-dimensional binning in $x$, $Q^2$, $z$ and $P_T$ consisting of nominally 12 x 8 x 12 x 12 bins has been selected where typically bins are combined for displaying purposes but for global fits this fine binning is used directly. The bin boundaries are given in Table \ref{tab:binning}. The variable $z$, which is roughly the momentum fraction of the struck parton a hadron carries is defined as:
\begin{equation}
    z = \frac{p\cdot P_h}{p\cdot q} \quad,
\end{equation} 
where $p$ is the four-momentum of the incoming nucleon, $P_h$ that of the detected hadron and $q$ is the momentum transfer. 
$P_T$ is the transverse momentum of the final-state hadron relative to the virtual photon direction in the frame where the incoming nucleon is at rest. 

Additionally, in each kinematic bin, the events are put into a two-dimensional histogram in the two azimuthal angles $\phi_S$ and $\phi_h$ with 16 equidistant bins in each dimension. Keeping both dimensions separated instead of directly using the angular combinations relevant to one particular azimuthal moment reduces the amount of uncertainties that can be introduced by smearing and reconstruction effects. 

\begin{figure}
    \centering
    \includegraphics[trim={2cm 17cm 8cm 4.8cm},clip,width=0.45\textwidth]{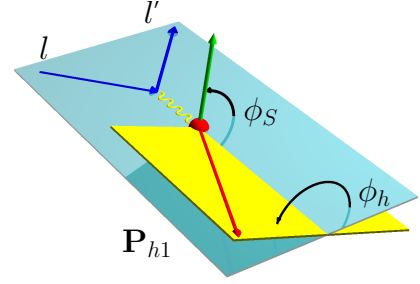}
    \caption{Azimuthal angles $\phi_h$ of the final state hadron ($\mathbf{P_h}$, red arrow) and $\phi_S$ of the transverse nucleon spin  direction (green arrow) in the lepton scattering plane around the virtual photon.}
    \label{fig:sidis}
\end{figure}

The two azimuthal angles are again defined around the virtual photon axis in the nucleon rest frame between the lepton scattering plane and either the transverse momentum of the final-state hadron ($\phi_h$) or the transverse spin direction of the incoming nucleon ($\phi_S$), as shown in Fig.~\ref{fig:sidis}.

\begin{table}[htb]
    \centering
    \begin{tabular}{c|c}
    Kinematic variable & Bin boundaries \\ \hline  & \\ 
        $x$ & $1.0x10^{-4}$, $2.154x10^{-4}$, $4.641x10^{-4}$, \\ &
					       $1.0x10^{-3}$,  $2.154x10^{-3}$,  $4.641x10^{-3}$,\\ &
					       $1.0x10^{-2}$, $2.154x10^{-2}$, $4.641x10^{-2}$,\\ &
					       $1.0x10^{-1}$, $2.154x10^{-1}$, $4.641x10^{-1}$,\\ &$1.0x10^{0}$ \\ &
					       \\ \hline   & \\ 
        $Q^2$ &  $1.0x10^{0}$, $3.162x10^{0}$,\\ &$1.0x10^1$, $3.162x10^1$,\\ &$1.0x10^2$, $3.162x10^2$,\\ &$1.0x10^3$, $3.162x10^3$,\\ &$1.0x10^4$\\ &
        \\ \hline  & \\ 
        $z$  &   0, 0.05, 0.1, 0.15, 0.2, 0.3, \\ &0.4, 0.5, 0.6, 0.7, 0.8, 0.9, 1.0 \\ & 
        \\ \hline  & \\
        $P_T$ & 0, 0.05, 0.1, 0.2, 0.3, 0.5, \\ &0.7, 0.9, 1.2, 1.5, 1.8, 2.4 ,4.0 \\ & \\ \hline
    \end{tabular}
    \caption{Kinematic bin boundaries in the main 4-dimensional binning used for the single spin asymmetry evaluation.}
    \label{tab:binning}
\end{table}

\section{Asymmetry reweighting}

All suitable events are reweighted based on the global fits of the SIDIS and $e^+e^-$ data that were extracted by the Torino group \cite{Anselmino:2008jk,Anselmino:2008sga} for Sivers and Collins asymmetries, respectively.

The weights were generated based on the true $x$, $Q^2$, $z$, $P_T$, $\phi_S$, $\phi_h$ and the true parton flavors and hadron IDs. 
In the provided code (see \cite{alexeigit}), the structure functions for both Sivers moment (proportional to a $\sin(\phi_h-\phi_S)$ moment), Collins moment (proportional to a $\sin(\phi_h+\phi_S)$ moment) and unpolarized TMDs were extracted for each event to obtain that event's weight when filling the reconstructed distributions. As this global fit is a leading order fit, the momenta, kinematics and flavors of the MC could be directly applied. 
Last, spin effects were applied by randomly assigning events to either spin up or spin down proton states where the spin-down weights had an additional phase of $\pi$ added to invert the sign of each moment in accordance with the proton spin pointing downward instead of upward.

At the time of this global fit, not all parton flavors and hadron types were included. Generally sea quark flavors were not present in either polarized structure functions (and hence the asymmetry weights were set to unity), as were the gluons for the Sivers function (and trivially not present for transversity). Furthermore also kaons were not implemented for Collins asymmetries so also their weights were identical to unity. 

As all parameterizations were based on the unpolarized distribution or fragmentation counter parts, the same unpolarized functions had to be used in the re-weighting as well, namely GRV98 \cite{Gluck:1998xa} for the distributions and DSS07\cite{deFlorian:2007aj} for the fragmentation functions. Those enter directly when calculating the polarized and unpolarized structure functions discussed above. 
\section{Reconstructed Asymmetries in relation to detector effects}
\subsection{Asymmetry extraction}
In each kinematic bin (or for projections after combining various bins) the single spin asymmetries are calculated similar to a real experiment using the formula:
\begin{equation}
    A_{UT}(\phi_S,\phi_h) =  \frac{N^+(\phi_S,\phi_h)-N^-(\phi_S,\phi_h)}{N^+(\phi_S,\phi_h)+N^-(\phi_S,\phi_h)}\quad,
\end{equation}
where $\pm$ indicate the two artificially created spin states and $N$ corresponds to the number of events in that state, kinematic and angular bin. This procedure is performed for both events in all the true kinematics as well as all the reconstructed kinematics. 

The corresponding Sivers and Collins asymmetries are then fit simultaneously in a two-dimensional fit that is given by:
\begin{equation}
    F(\phi_S,\phi_h) = A_{COL}\sin(\phi_h+\phi_S) + A_{SIV}\sin(\phi_h-\phi_S)\quad.
\end{equation}
It is important to fit both moments simultaneously in a two-dimensional fit instead of performing one-dimensional fits in the projections as those may in principle suffer when the acceptance is not perfect. 

\begin{figure*}
    \centering
    \includegraphics[width=0.9\textwidth]{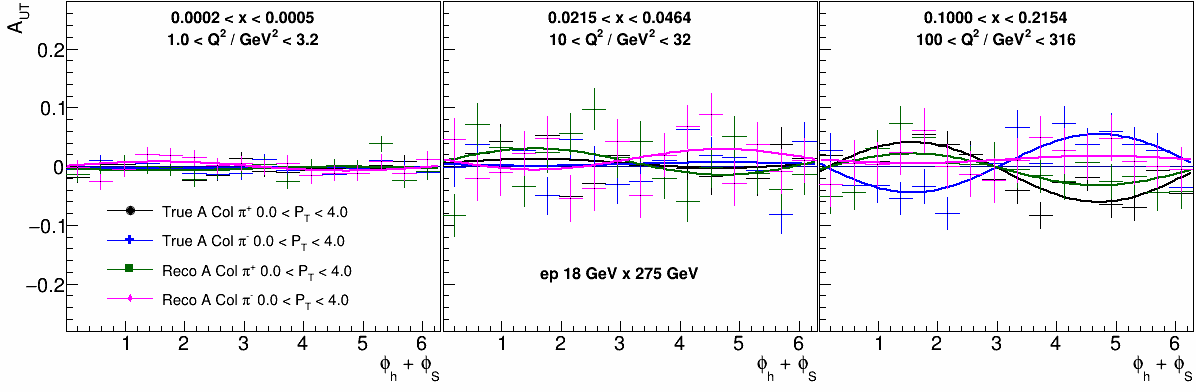}
    \includegraphics[width=0.9\textwidth]{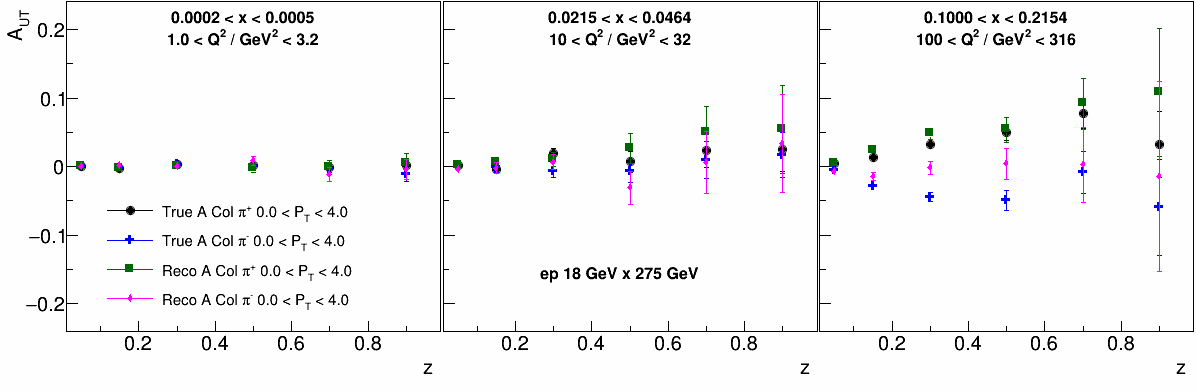}
    \includegraphics[width=0.9\textwidth]{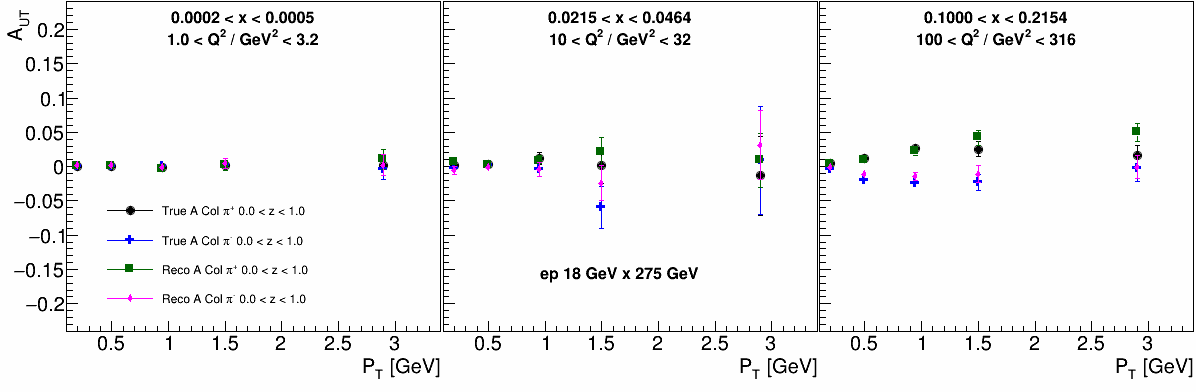}
    \caption{Top: One-dimensional azimuthal asymmetry projection for the Collins type modulation as a function of the angular combination $\phi_h+\phi_S$ are shown for the specified $x$, $Q^2$ bins, integrated over transverse momentum and in an intermediate $z$ bin for 18x275 GeV collisions. True and reconstructed data shown for positive pions (black and green respectively) and negative pions (blue and magenta, respectively). The fits to the sine modulations are displayed as well in the corresponding colors. 
    Middle: Extracted Collins-type asymmetries in the same $x$ and $Q^2$ bins, integrated over $P_T$ and displayed as a function of $z$.
    Bottom: Extracted Collins-type asymmetries in the same $x$ and $Q^2$ bins, integrated over $z$ and displayed as a function of $P_T$.}
    \label{fig:sampleasy}
\end{figure*}

\begin{figure*}
    \centering
    \includegraphics[width=0.9\textwidth]{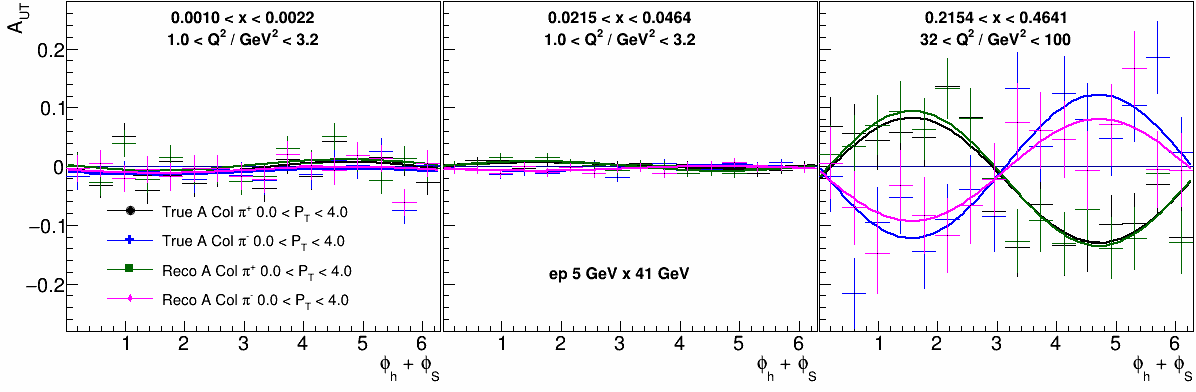}
    \includegraphics[width=0.9\textwidth]{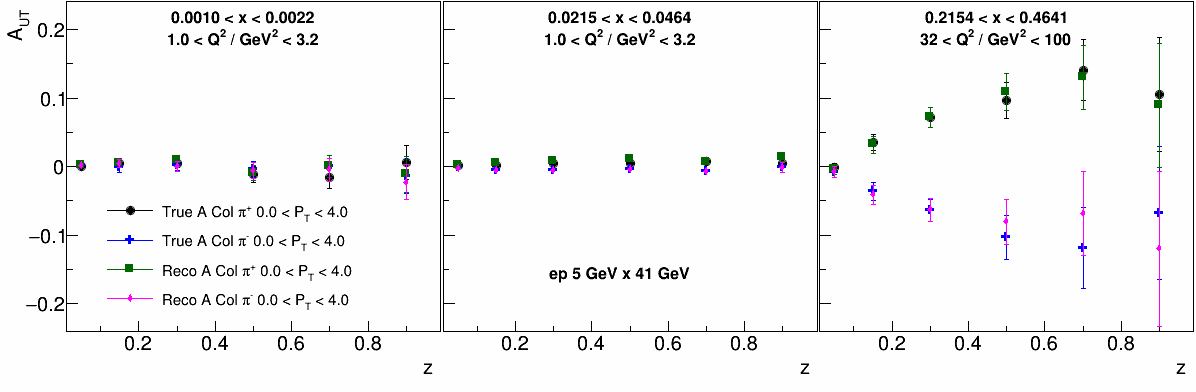}
    \includegraphics[width=0.9\textwidth]{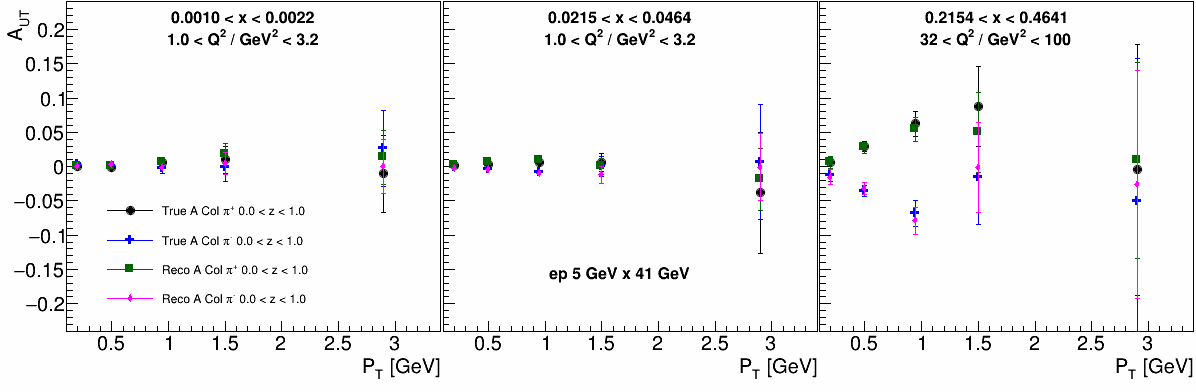}
    \caption{Top: One-dimensional azimuthal asymmetry projection for the Collins type modulation as a function of the angular combination $\phi_h+\phi_S$ are shown for the specified $x$, $Q^2$ bins, integrated over transverse momentum and in an intermediate $z$ bin for 5x41 GeV collisions. True and reconstructed data shown for positive pions (black and green respectively) and negative pions (blue and magenta, respectively). The fits to the sine modulations are displayed as well in the corresponding colors. 
    Middle: Extracted Collins-type asymmetries in the same $x$ and $Q^2$ bins, integrated over $P_T$ and displayed as a function of $z$.
    Bottom: Extracted Collins-type asymmetries in the same $x$ and $Q^2$ bins, integrated over $z$ and displayed as a function of $P_T$.}
    \label{fig:sampleasy2}
\end{figure*}

However, examples of these one-dimensionally projected results are displayed in selected kinematic bins in Fig.~\ref{fig:sampleasy} for the Sivers moments, and in Fig.~\ref{fig:sampleasy2} for the Collins moments. These figures are only for visualization purposes since they use these simplified one-dimensional azimuthal binnings. Their fits are displayed as well in that figure, but it should be noted, that those are not directly used for the physics results and impact studies as these one-dimensional projections are more prone to mis-reconstruct the actual asymmetries.

\subsection{Reconstructed asymmetries}
\begin{figure*}
    \centering
    \includegraphics[width=0.8\textwidth]{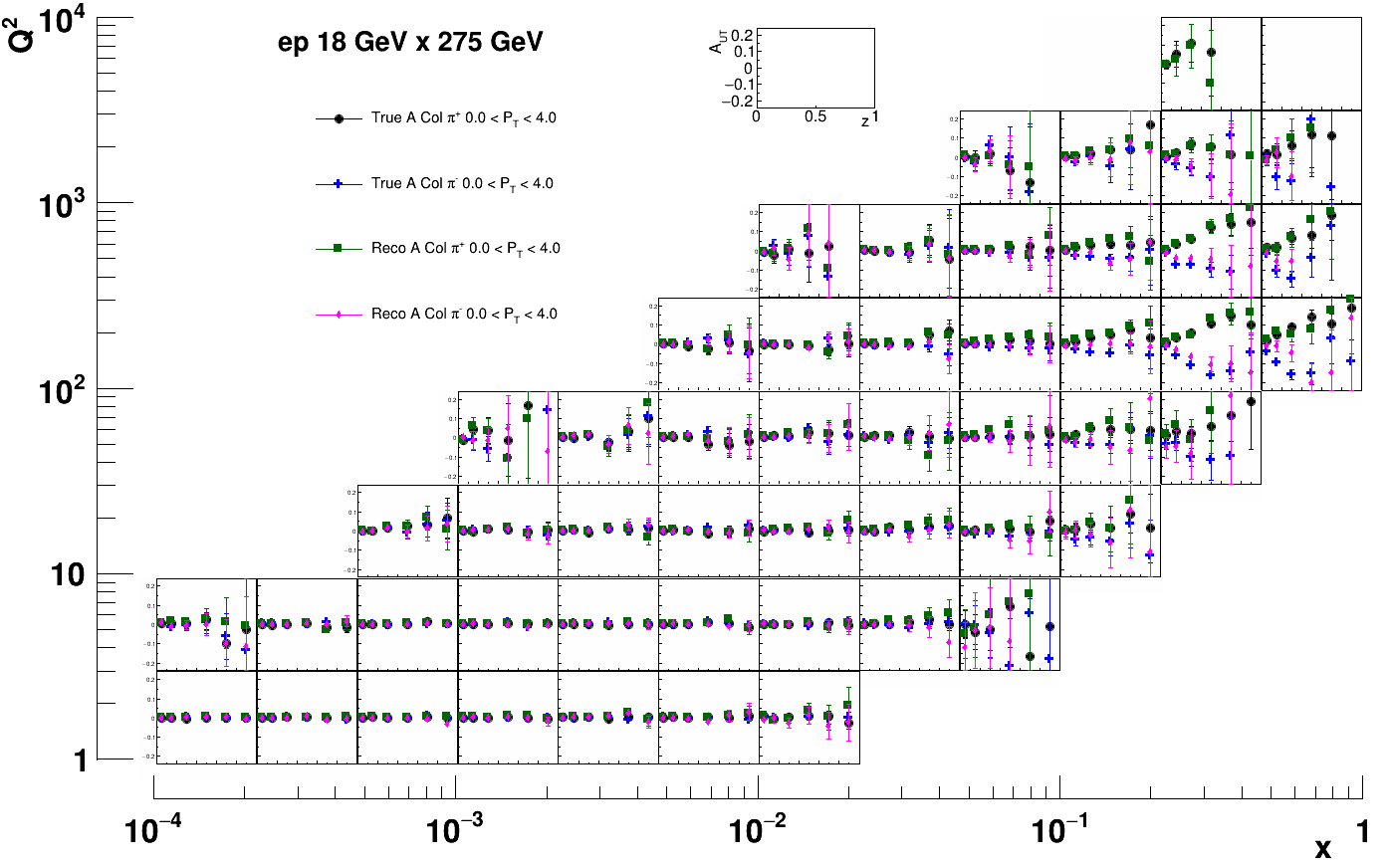}
    \caption{Collins asymmetries as a function of $z$ in bins of $x$ and $Q^2$ for 18 GeV electrons on 275 GeV protons for positive (black, green) and negative (blue, purple) pions. The asymmetries in the true kinematics are shown in black and blue symbols while the asymmetries in the reconstructed kinematics are shown in blue and purple symbols.}
    \label{fig:reco_dat1_col}
\end{figure*}

\begin{figure*}
    \centering
    \includegraphics[width=0.8\textwidth]{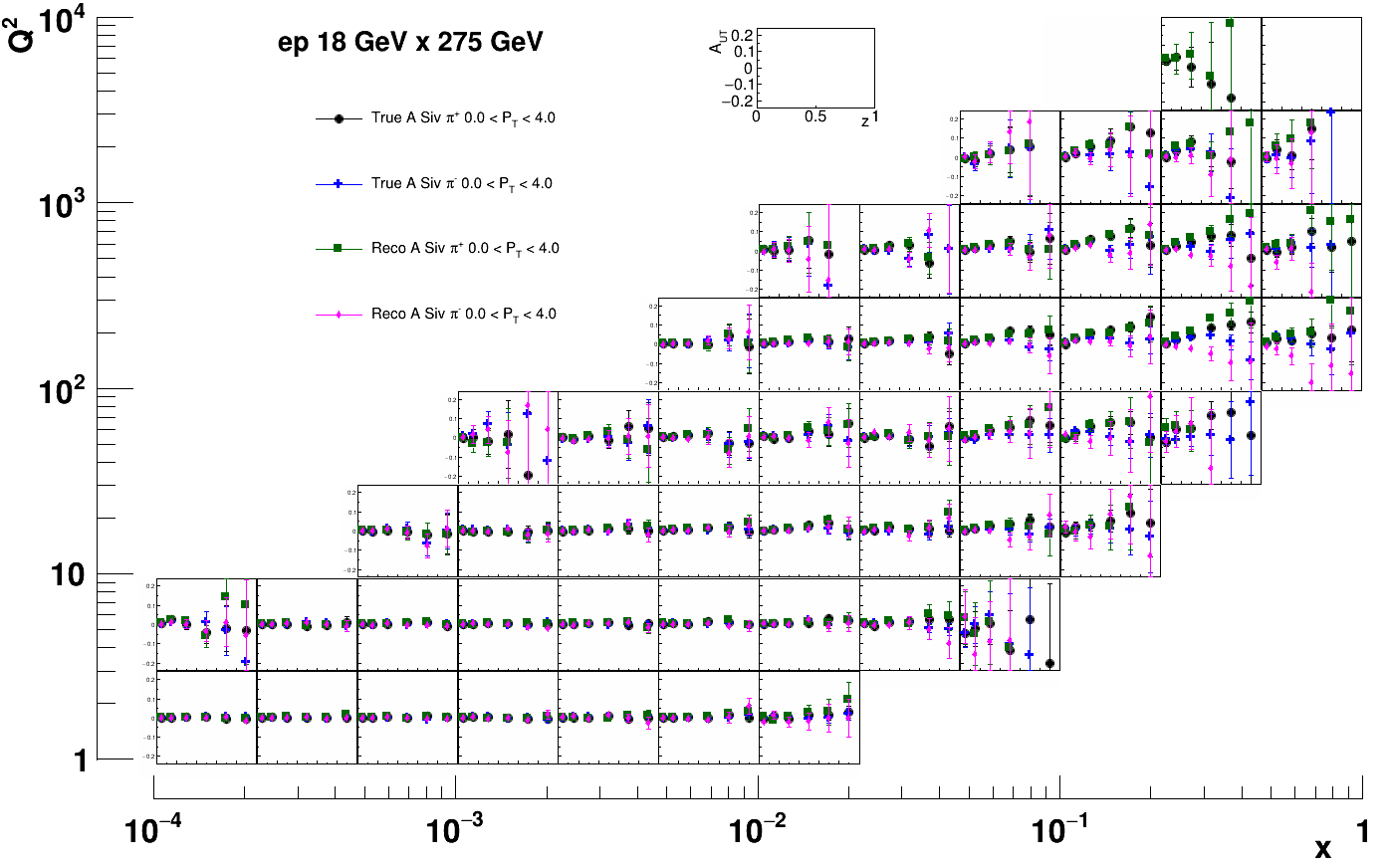}
    \caption{Sivers asymmetries as a function of $z$ in bins of $x$ and $Q^2$ for 18 GeV electrons on 275 GeV protons for positive (black, green) and negative (blue, purple) pions. The asymmetries in the true kinematics are shown in black and blue symbols while the asymmetries in the reconstructed kinematics are shown in blue and purple symbols.}
    \label{fig:reco_dat1_siv}
\end{figure*}

\begin{figure*}
    \centering
    \includegraphics[width=0.8\textwidth]{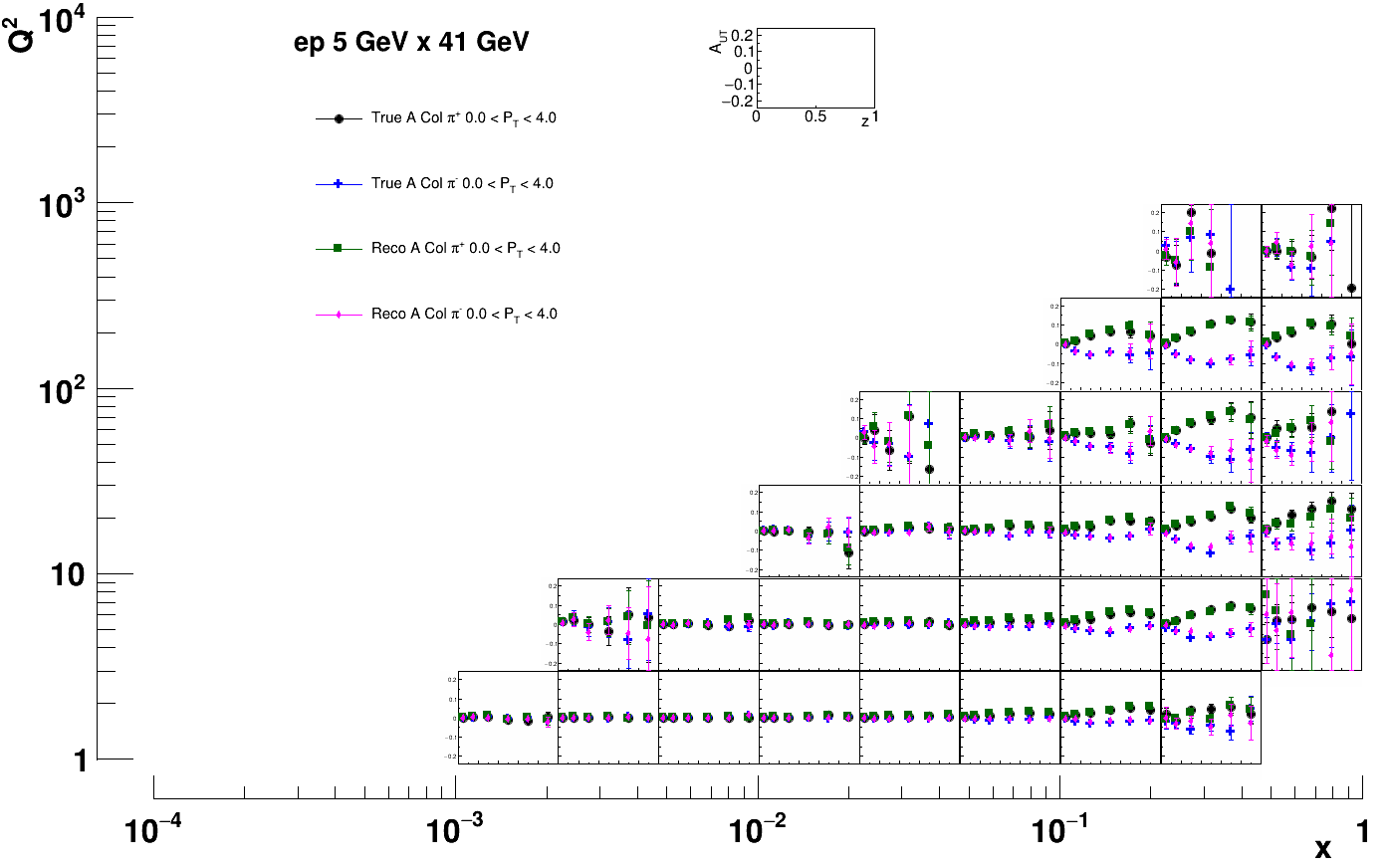}
    \caption{Collins asymmetries as a function of $z$ in bins of $x$ and $Q^2$ for 5 GeV electrons on 41 GeV protons for positive (black, green) and negative (blue, purple) pions. The asymmetries in the true kinematics are shown in black and blue symbols while the asymmetries in the reconstructed kinematics are shown in blue and purple symbols.}
    \label{fig:reco_dat16_col}
\end{figure*}

\begin{figure*}
    \centering
    \includegraphics[width=0.8\textwidth]{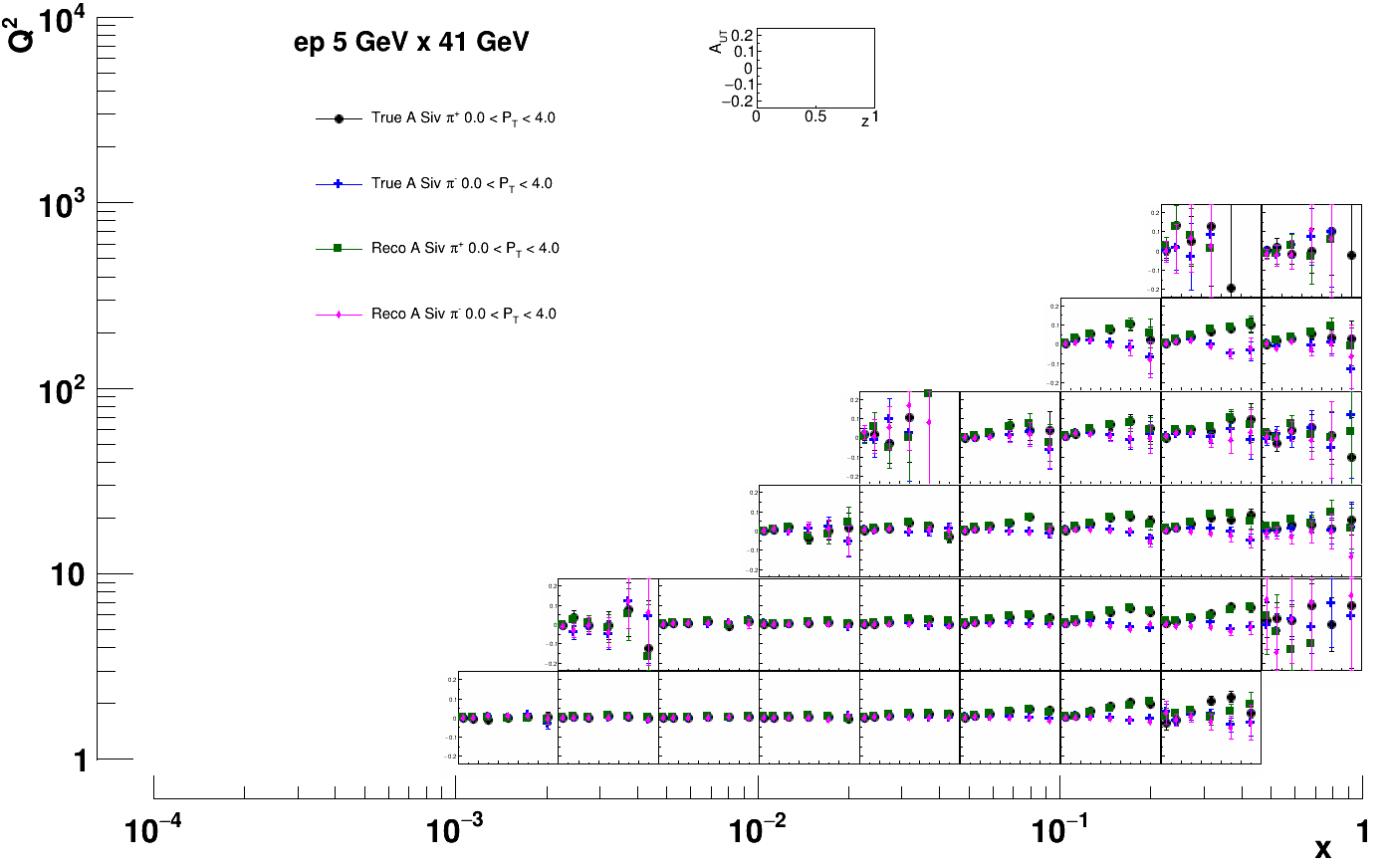}
    \caption{Sivers asymmetries as a function of $z$ in bins of $x$ and $Q^2$ for 5 GeV electrons on 41 GeV protons for positive (black, green) and negative (blue, purple) pions. The asymmetries in the true kinematics are shown in black and blue symbols while the asymmetries in the reconstructed kinematics are shown in blue and purple symbols.}
    \label{fig:reco_dat16_siv}
\end{figure*}

The reconstructed asymmetries are shown in various projections of fractional energy and transverse momentum for the true and the reconstructed values. One can see the projections for charged pions where two $z$ bins and all transverse momentum bins were combined in Fig.~\ref{fig:reco_dat1_col} and Fig.~\ref{fig:reco_dat1_siv} for the highest collision energies and similarly in Figs.~\ref{fig:reco_dat16_col} and \ref{fig:reco_dat16_siv} for the lowest energies. At both energies one can clearly see the nonzero asymmetries at higher $x$ and the different signs for positive and negative Collins asymmetries that arise from the different signs of the favored and disfavored Collins functions, together with the different signs of up and down quark transversity distributions. 
At lower $x$ all asymmetries are consistent with zero as the parameterizations so far lack the sea quark TMDs and also the valence quark TMDs are parametrized to be small. 

In the Sivers asymmetries the positive pions are again showing positive asymmetries at larger $x$, in accordance with the fixed target SIDIS results. In the case of negative pions, however, the asymmetries are generally small, even at higher $x$ which originates in cancellations between the contributions from up quark Sivers function with disfavored fragmentation and down quark Sivers function with favored fragmentation as up and down quark Sivers functions are opposite in sign.

In all the regions where nonzero asymmetries can be seen, one can also observe that generated and reconstructed asymmetries agree generally very well. This shows that the amount of smearing in all of the kinematic variables is moderate enough that the asymmetries can be reasonably well extracted. One can also see, that despite the generated luminosities being substantially below the expected luminosities, the statistical uncertainties are overall small, allowing to use a finer binning in the various kinematic variables.

\section{Projected asymmetries and estimate of systematic uncertainties}
After extracting the weighted asymmetries in the given statistics, a next step is to extrapolate these measurements to the full statistics that are aimed for at the EIC. 
In this study it was assumed that at each collision energy an accumulated luminosity of 10 fb$^{-1}$ can be obtained and in the projection figures all statistical uncertainties were scaled to this luminosity. A 70\% beam polarization was assumed which scales all statistical uncertainties with the inverse of this polarization. For the systematic uncertainties a similar approach as in the EIC yellow report has been chosen, where as a conservative estimate the differences between true and reconstructed asymmetries based on the re-weighting were assigned. 
This estimate is for the most part clearly too conservative as in reality an unfolding of the asymmetries would be performed. However, such an unfolding would require substantially more, and preferably independent, MC simulations as well as a much more detailed description of all detector components, etc. Even in an unfolding procedure, if the discrepancies between true and reconstructed asymmetries get large (i.e. larger off-diagonal elements in the smearing matrix), also the uncertainties in the unfolding will increase and therefore taking at this point these differences as uncertainties is a prudent, albeit very conservative approach. 
Not shown in the figures or data tables that are provided to the theorists for impact evaluations is an additional global 3\% relative systematic uncertainty related to the precision of extracting the beam polarization in the EIC.  

\begin{figure*}
    \centering
    \includegraphics[width=0.95\textwidth]{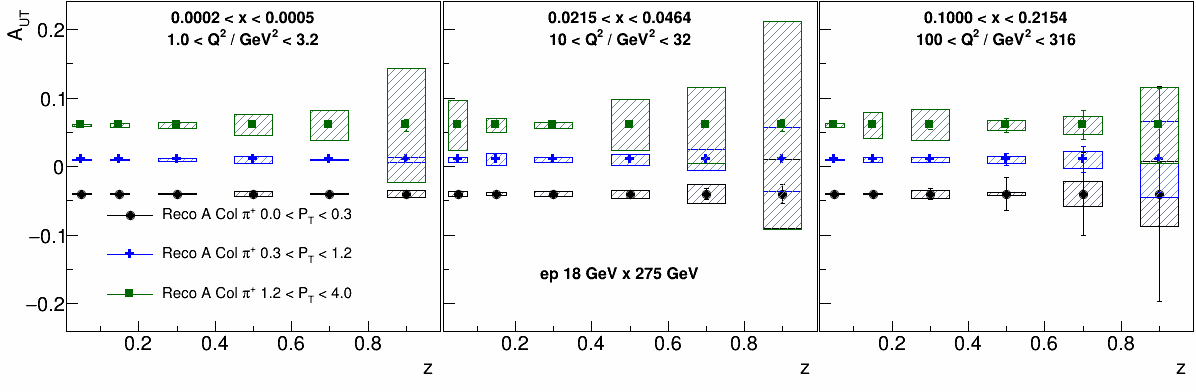}
    \includegraphics[width=0.95\textwidth]{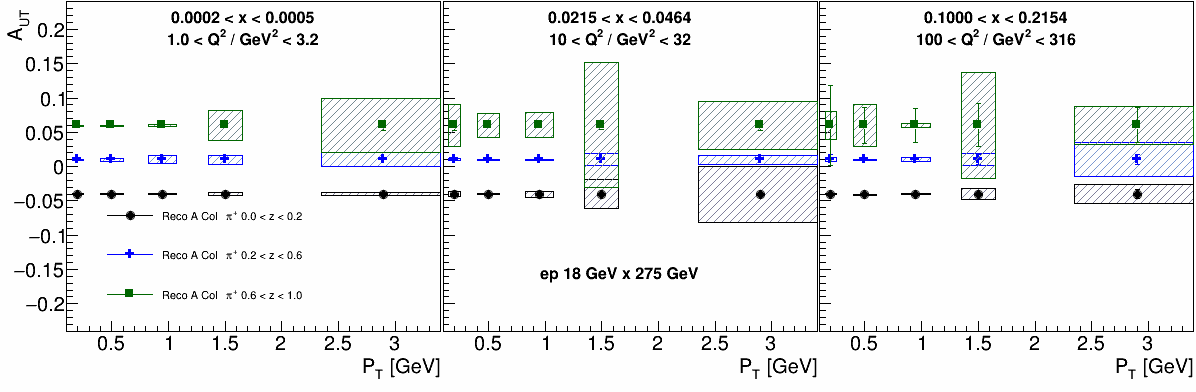}
    \caption{Projected $\pi^+$ Collins asymmetry statistical and systematic uncertainties as a function of either $z$ (top panel) in bins of $P_T$ or as a function of $P_T$ in bins of $z$ (bottom panel) for three select $x$ and $Q^2$ bins. The asymmetries are shown at arbitrary values for better visibility. The statistical uncertainties are extrapolated to an accumulated luminosity of 10 fb$^{-1}$ for the 18 GeV x 275 GeV energy option. For better visibility either 4 bins in $P_T$ and 2 bins in $z$ were combined or vice versa.}
    \label{fig:projectionsample_Col_18x275}
\end{figure*}

\begin{figure*}
    \centering
    \includegraphics[width=0.95\textwidth]{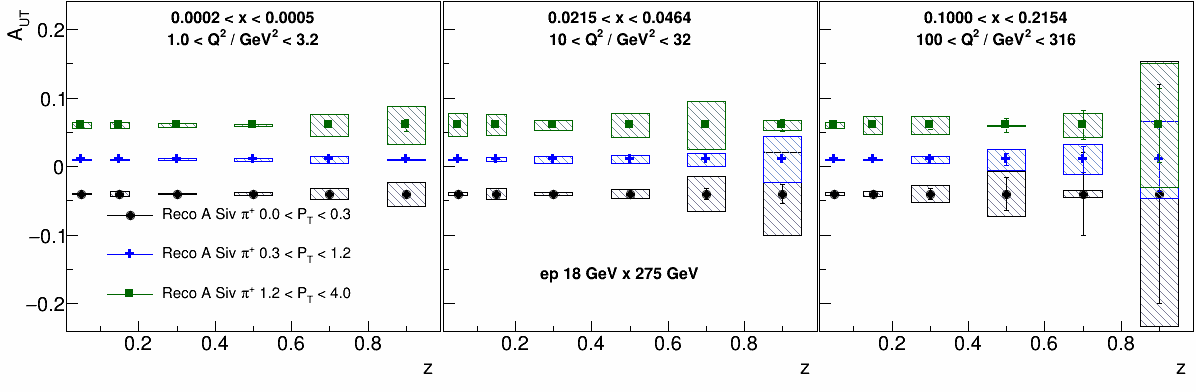}
    \includegraphics[width=0.95\textwidth]{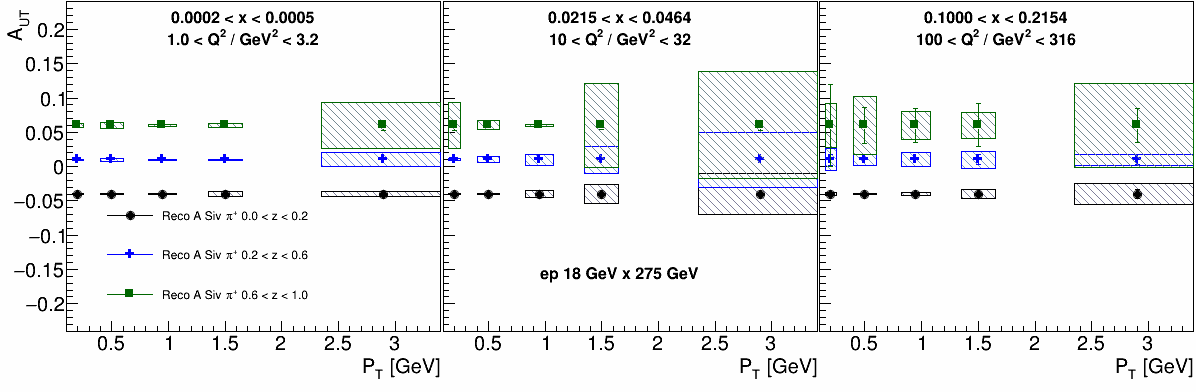}
    \caption{Projected $\pi^+$ Sivers asymmetry statistical and systematic uncertainties as a function of either $z$ (top panel) in bins of $P_T$ or as a function of $P_T$ in bins of $z$ (bottom panel) for three select $x$ and $Q^2$ bins. The asymmetries are shown at arbitrary values for better visibility. The statistical uncertainties are extrapolated to an accumulated luminosity of 10 fb$^{-1}$ for the 18 GeV x 275 GeV energy option. For better visibility either 4 bins in $P_T$ and 2 bins in $z$ were combined or vice versa.}
    \label{fig:projectionsample_Siv_18x275}
\end{figure*}

An example of the expected uncertainties in three select $x$ and $Q^2$ bins are shown in Figs.\ref{fig:projectionsample_Col_18x275} and \ref{fig:projectionsample_Siv_18x275} for the highest collision energies for Collins and Sivers asymmetries, respectively. 
Already from this simple figure one can take away several important aspects of the expected uncertainties using the EIC and in particular ECCE. At low values of $x$ and $Q^2$ the cross sections are largest and therefore the statistical uncertainties are generally very small; however, as the asymmetries are also expected to be well below the 1 \% level there, such precision is quite necessary. One also sees that at combinations of higher transverse momenta or fractional energies the systematic uncertainties are increasing which is likely related to the smearing in these two kinematic variables, given that the azimuthal angles do not show larger smearing effects. 
This behavior is even more prominent when going toward higher $x$ and $Q^2$ where also the smearing in $x$ and $Q^2$ is becoming more relevant. In terms of statistical uncertainties the expected uncertainties stay below the one per-cent level as long as neither $z$ nor $P_T$ become too large.

\begin{figure*}
    \centering
    \includegraphics[width=0.95\textwidth]{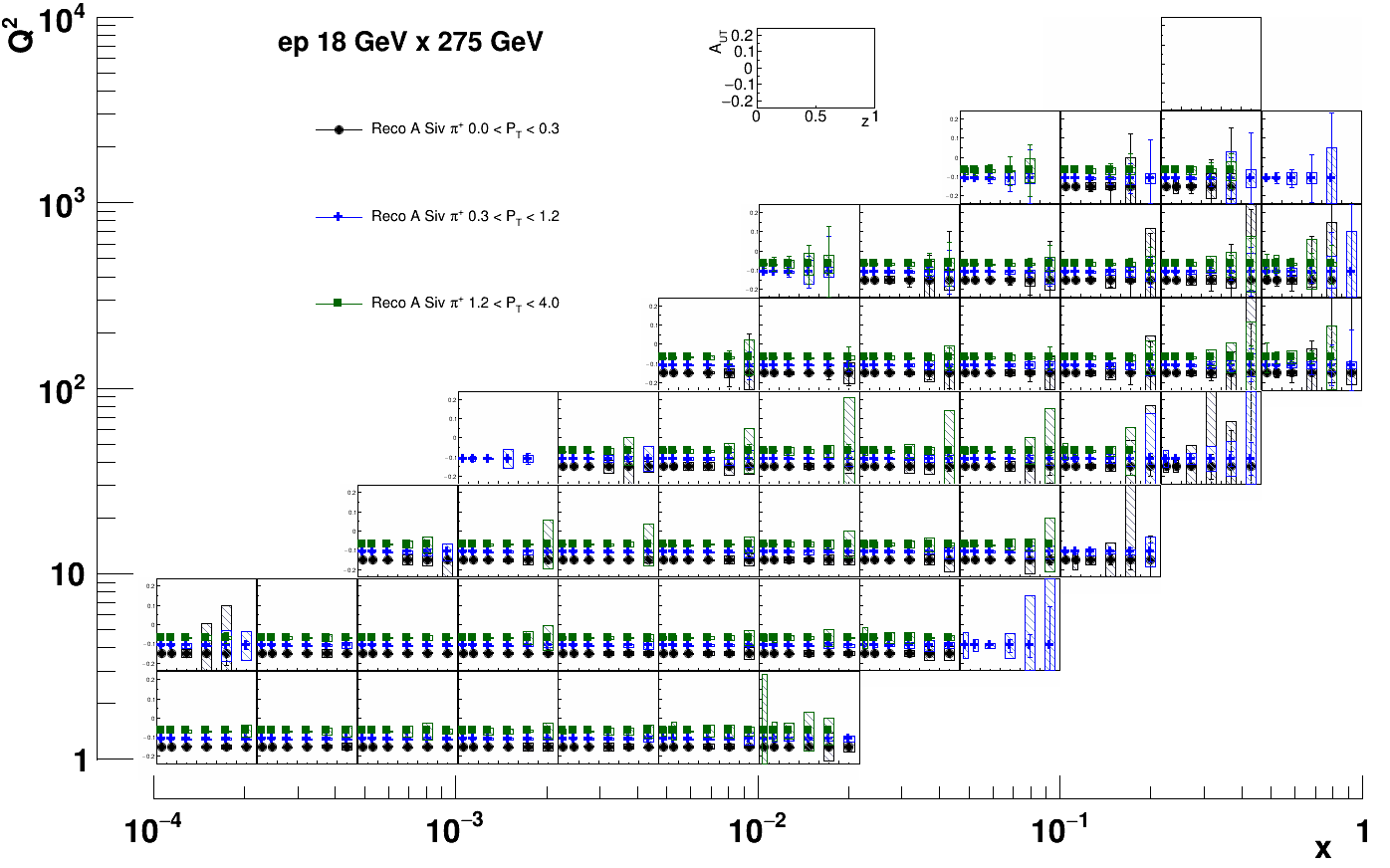}
    \caption{Projected $\pi^+$ Sivers asymmetry statistical and systematic uncertainties as a function of $z$ in bins of $P_T$, $x$ and $Q^2$ shown at arbitrary asymmetry values for better visibility. The statistical uncertainties are extrapolated to an accumulated luminosity of 10 fb$^{-1}$ for the 18 GeV x 275 GeV energy option.}
    \label{fig:projections_Siv_18x275}
\end{figure*}

\begin{figure*}
    \centering
    \includegraphics[width=0.95\textwidth]{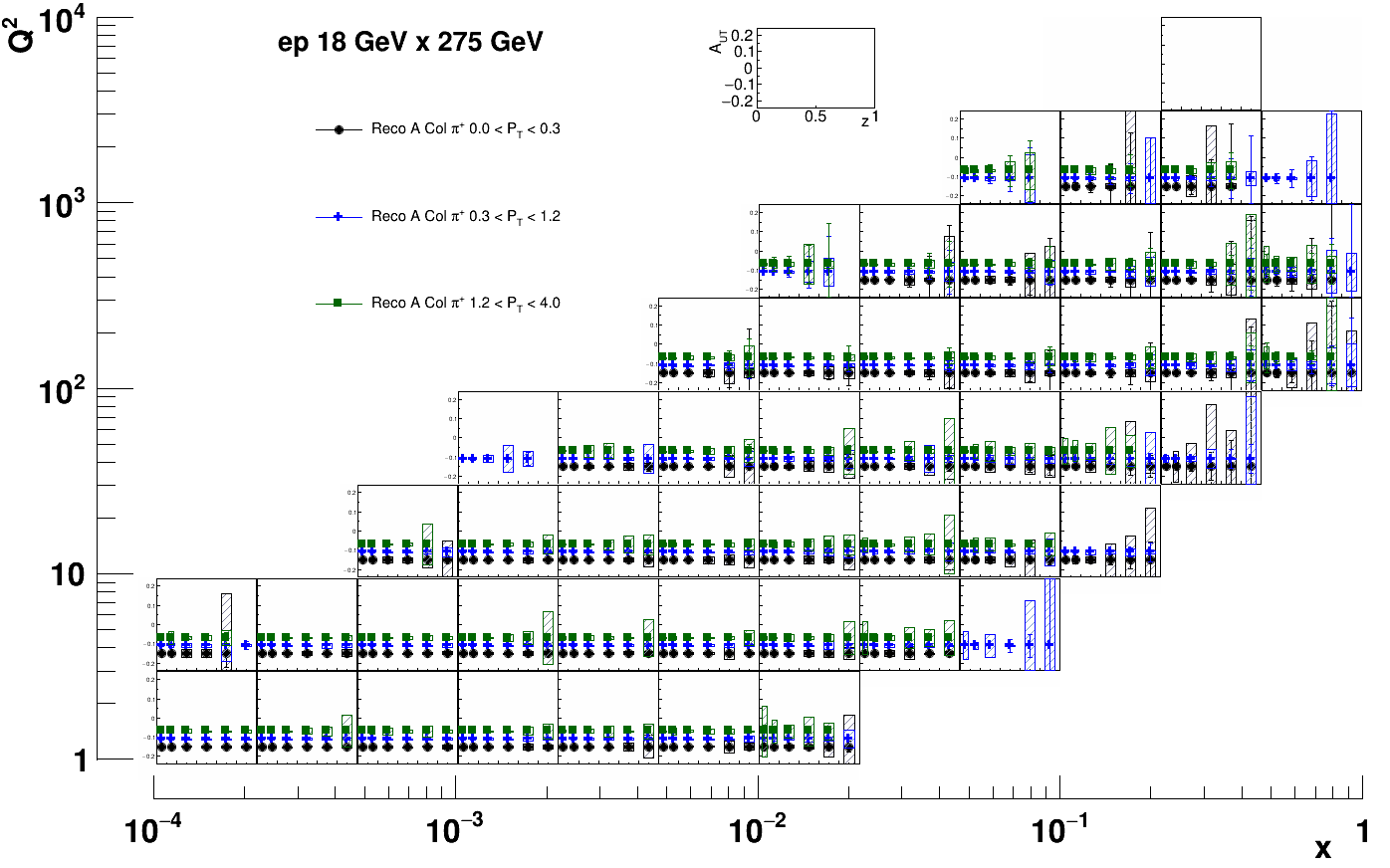}
    \caption{Projected $\pi^+$ Collins asymmetry statistical and systematic uncertainties as a function of $z$ in bins of $P_T$, $x$ and $Q^2$ shown at arbitrary asymmetry values for better visibility. The statistical uncertainties are extrapolated to an accumulated luminosity of 10 fb$^{-1}$ for the 18 GeV x 275 GeV energy option.}
    \label{fig:projections_Col_18x275}
\end{figure*}

A simplified display of the expected uncertainties for all $x$ and $Q^2$ bins are shown in Fig.~\ref{fig:projections_Siv_18x275} for the Sivers asymmetries at the highest collision energy and in Fig.~\ref{fig:projections_Col_18x275} for the Collins asymmetries where for visibility two $z$ bins and four $P_T$ bins were combined. As can be seen, a sub percent level of precision can be reached over a large range of $x$ and $Q^2$ bins, with this collision energy mostly covering the lower $x$ and the higher $Q^2$ bins best. Following the $z$ dependence of fragmentation functions the statistical and systematic uncertainties tend to increase toward higher $z$ while also the uncertainties at higher transverse momenta are generally the largest likely following similar reasons.

\begin{figure*}
    \centering
    \includegraphics[width=0.95\textwidth]{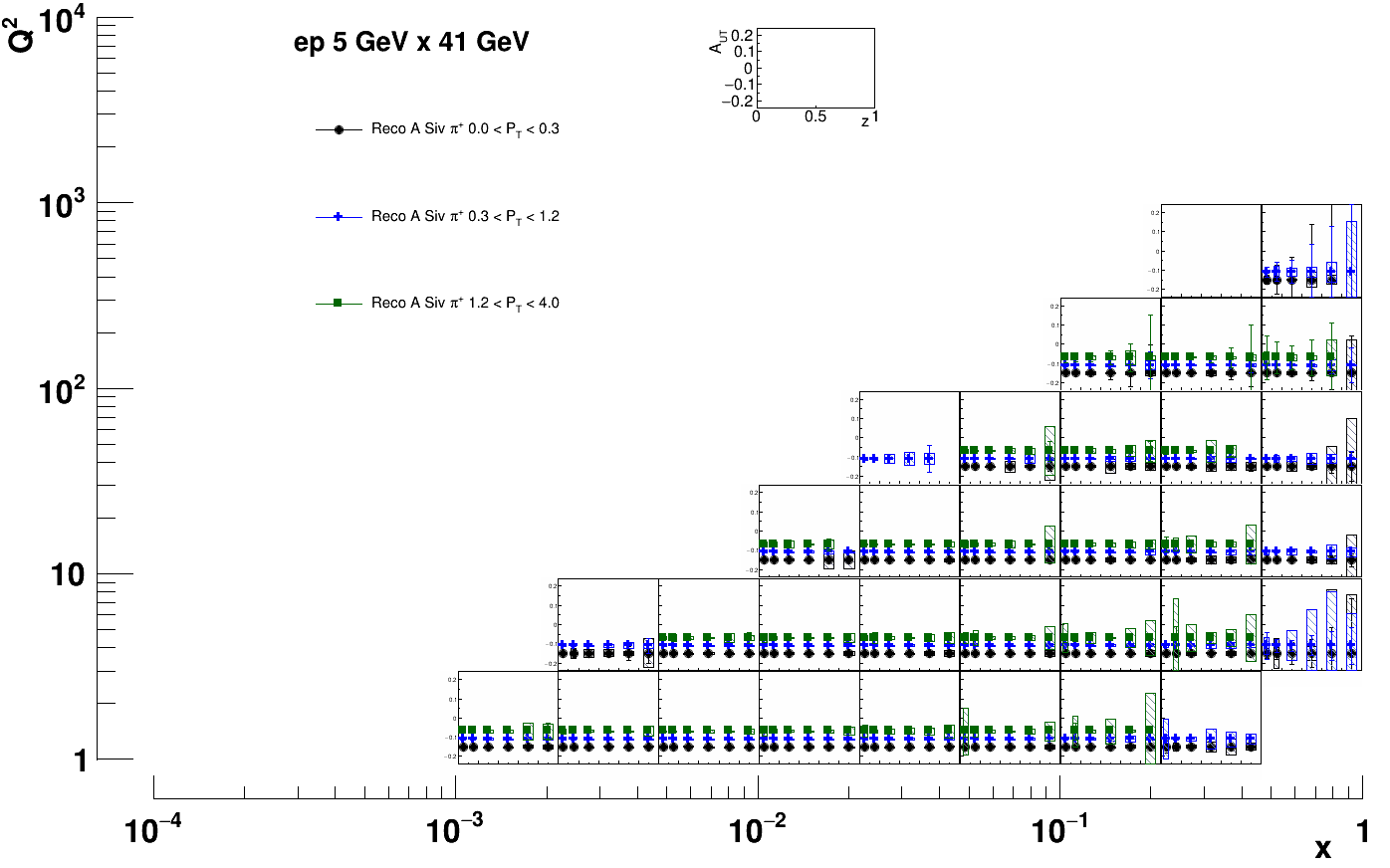}
    \caption{Projected $\pi^+$ Sivers asymmetry statistical and systematic uncertainties as a function of $z$ in bins of $P_T$, $x$ and $Q^2$ shown at arbitrary asymmetry values for better visibility. The statistical uncertainties are extrapolated to an accumulated luminosity of 10 fb$^{-1}$ for the 5 GeV x 41 GeV energy option.}
    \label{fig:projections_Siv_5x41}
\end{figure*}

\begin{figure*}
    \centering
    \includegraphics[width=0.95\textwidth]{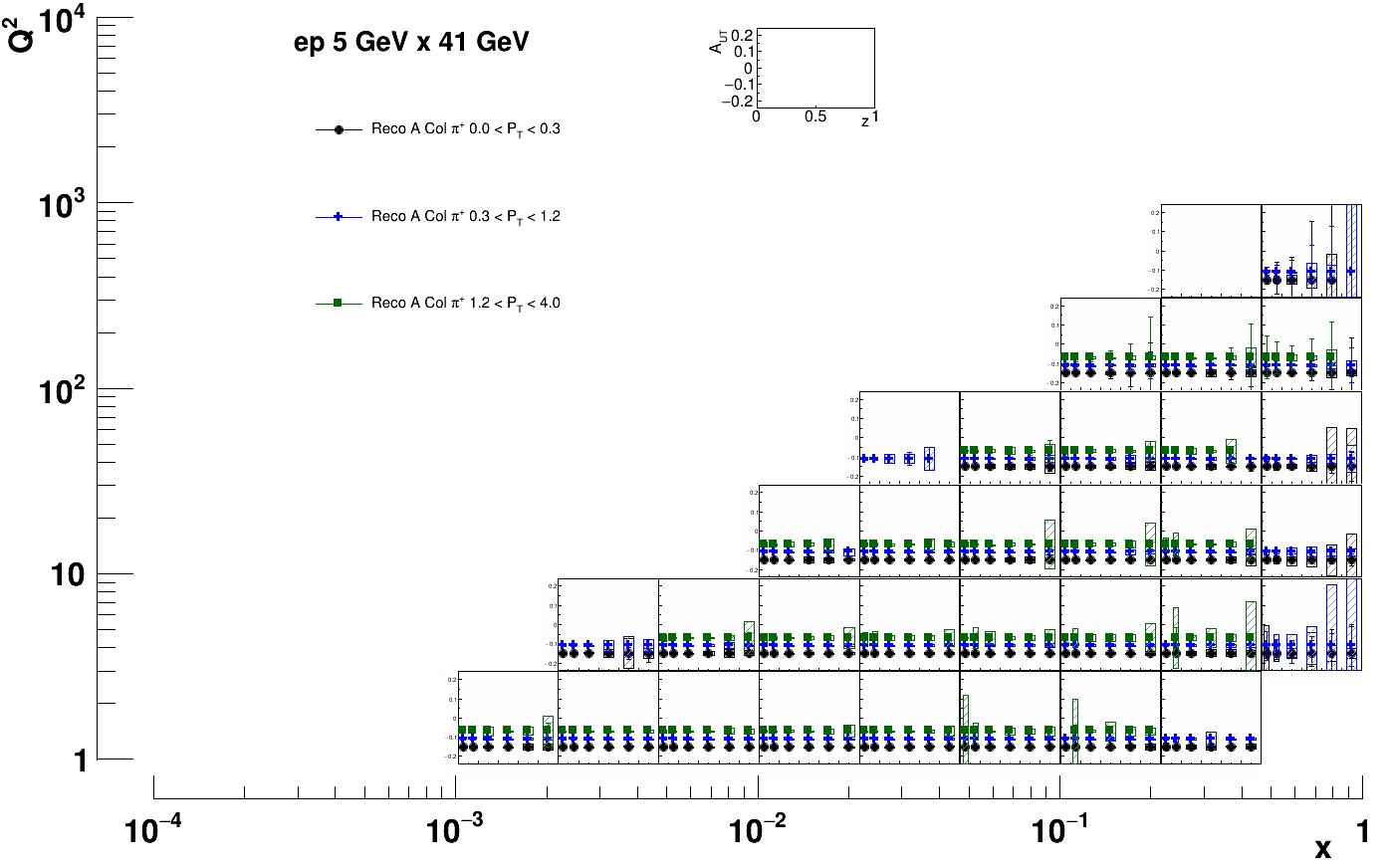}
    \caption{Projected $\pi^+$ Collins asymmetry statistical and systematic uncertainties as a function of $z$ in bins of $P_T$, $x$ and $Q^2$ shown at arbitrary asymmetry values for better visibility. The statistical uncertainties are extrapolated to an accumulated luminosity of 10 fb$^{-1}$ for the 5 GeV x 41 GeV energy option.}
    \label{fig:projections_Col_5x41}
\end{figure*}

The lower collision energies move successively further toward the bottom right of the $x$-$Q^2$ plane with the lowest collision energy covering the highest $x$ and lowest $Q^2$, as can be seen in Figs.~\ref{fig:projections_Siv_5x41} and \ref{fig:projections_Col_5x41} for Sivers and Collins asymmetries, respectively. Due to the generally lower center-of-mass energy, the phase space for larger transverse momenta is substantially reduced at higher $x$ and thus only precise measurements for transverse momenta below 1.2 GeV are possible there. However, the missing transverse momenta would anyway fall outside of the region that could be interpreted via TMDs \cite{Boglione:2019nwk}.

\begin{figure*}
    \centering
    \includegraphics[width=0.8\textwidth]{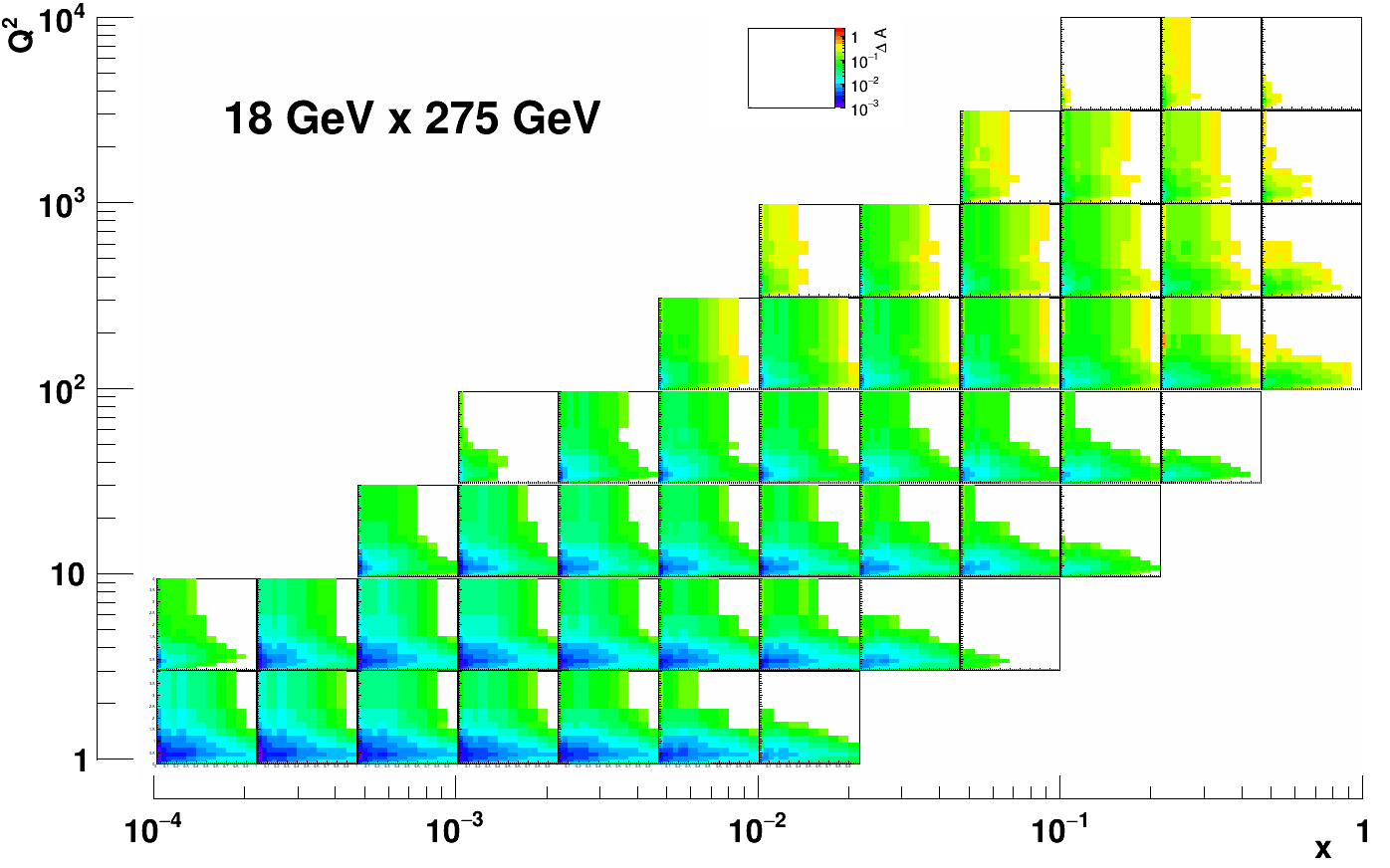}
    \caption{Projected $\pi^+$ Sivers asymmetry statistical uncertainties as a function of $z$ and $P_T$ in bins of $x$ and $Q^2$. The statistical uncertainties are extrapolated to an accumulated luminosity of 10 fb$^{-1}$ for the 18 GeV x 275 GeV energy option.}
    \label{fig:projectionsstat_Siv_18x275}
\end{figure*}

\begin{figure*}
    \centering
    \includegraphics[width=0.8\textwidth]{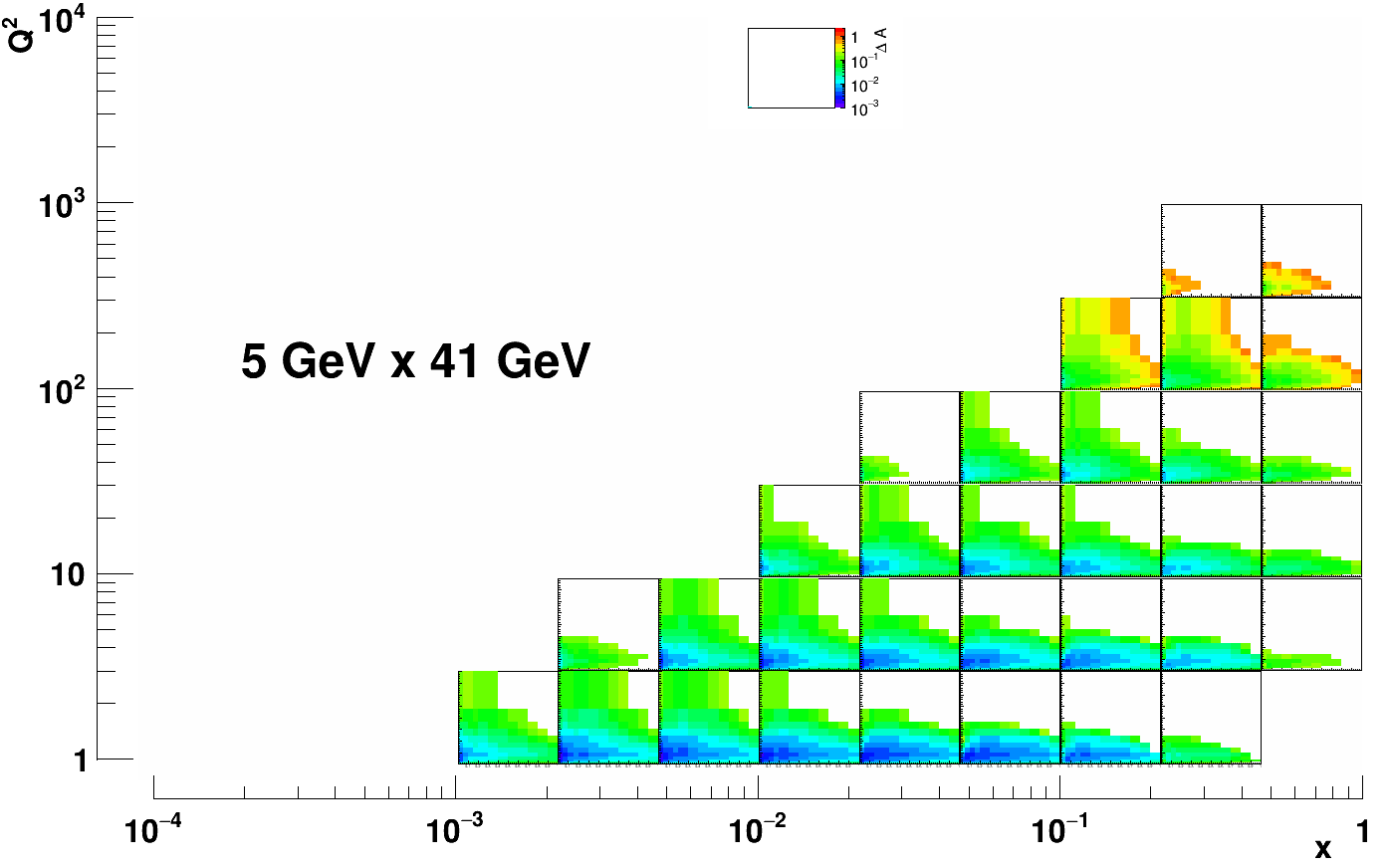}
    \caption{Projected $\pi^+$ Sivers asymmetry statistical uncertainties as a function of $z$ and $P_T$ in bins of $x$ and $Q^2$. The statistical uncertainties are extrapolated to an accumulated luminosity of 10 fb$^{-1}$ for the 5 GeV x 41 GeV energy option.}
    \label{fig:projectionsstat_Siv_5x41}
\end{figure*}

A simple visualisation of just the uncertainties for all 4-dimensional bins is shown in Fig.~\ref{fig:projectionsstat_Siv_18x275}. One can see that even in such a fine binning a sub per-cent level statistical precision can be reached, predominantly at lower $x$ and $Q^2$ and for lower transverse momenta and momentum fractions. However, over the whole phase-space the expected precision is still on the \% level except for the phase-space boundaries in $z$ and $P_T$. A similar estimate can be seen in Fig.~\ref{fig:projectionsstat_Siv_5x41} for the lowest collision energy. These figures also roughly display the phase space in $z$ and $P_T$ that can be covered at a given $x$ and $Q^2$ bin. In particular at higher $z$ and predominantly for higher $x$ and $Q^2$ the transverse momentum is rather limited. 

\subsection{Projected neutron pseudo-data}
As the {\sc geant} simulations require a large amount of computing time, full simulations of e+$^3$He collisions and the corresponding neutron information were not performed. Instead, the uncertainty ratios of the ECCE e+p simulated pseudo-data and that of the Yellow report pseudo-data were taken to scale the corresponding neutron data from the Yellow report. 
This also serves as a good gauge as to how similar the pseudo-data uncertainties, and hence the physics impact will be. It was found that the statistical uncertainties are on average a few per-cent larger in the full ECCE detector simulation compared to the parameterized simulations of the Yellow report. This is of course not surprising since both simulations cover the same rapidity regions. For pions they are even at times lower, as in the Yellow report the pseudo-data was cut as soon as the pion-kaon separation would be lower than three standard deviations, while in ECCE these regions are assumed to be included (though likely with increased systematic uncertainties due to the PID unfolding). 
In terms of the systematic uncertainties one does see an increase in the uncertainties by up to 20 \% which originates from the more realistic smearing in all the kinematic variables as the ECCE simulations do not only include all the realistic detector components but also support structure and detector material that impact the resolutions. However, it should be noted that these uncertainties are representing the worst-case scenario and in an actual measurement detector smearing would be appropriately unfolded.


\subsection{$Q^2$ dependence of asymmetries}

\begin{figure*}
    \centering
    \includegraphics[width=0.9\textwidth]{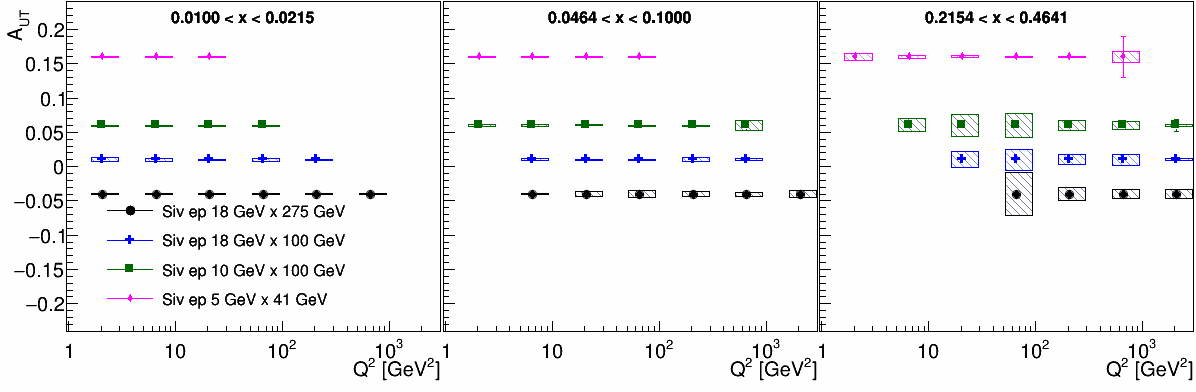}
    \caption{Example figure of the $Q^2$ dependence of Sivers asymmetries for $\pi^+$ for three $x$ bins after integrating over transverse and fractional momenta.}
    \label{fig:Sivevoe}
\end{figure*}
\begin{figure*}
    \centering
    \includegraphics[width=0.9\textwidth]{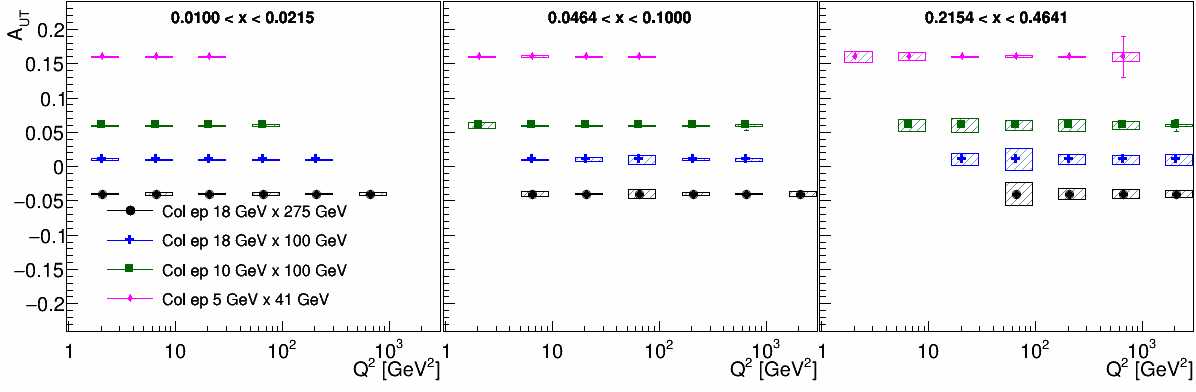}
    \caption{Example figure of the $Q^2$ dependence of Collins asymmetries for $\pi^+$ for three $x$ bins after integrating over transverse and fractional momenta.}
    \label{fig:Colevo}
\end{figure*}

Another important aspect of the transverse single spin asymmetries is the study of their scale dependence. For these, it is important to have as large a $Q^2$ lever arm at an $x$ value where the asymmetries are not too small at low scales. For some ranges of transverse momentum and fractional energies the expected asymmetries are extracted as a function of $Q^2$. Those are displayed in Fig.~\ref{fig:Sivevoe} for pion Sivers asymmetries, extrapolated to the 10 fb$^{-1}$ for each collision energy. One can see that at intermediate to higher $x$ a large range in $Q^2$ can be covered where the different collision energies aid each other in coverage. At lower scales the lower collision energies provide good coverage while at higher scales the higher collision energies provide coverage. Nevertheless, there is enough overlap such that systematic uncertainties from running periods with different energies can be cross checked. A similar figure for the Collins asymmetries can be seen in Fig.~\ref{fig:Colevo}.
This represents just an example of the sensitivities that can be obtained. In an actual analysis at least a rough binning of $P_T$ and $z$ would be kept as well as potentially choosing a finer binning in $Q^2$.

\begin{figure}[thb]
    \centering
    \includegraphics[width=0.48\textwidth]{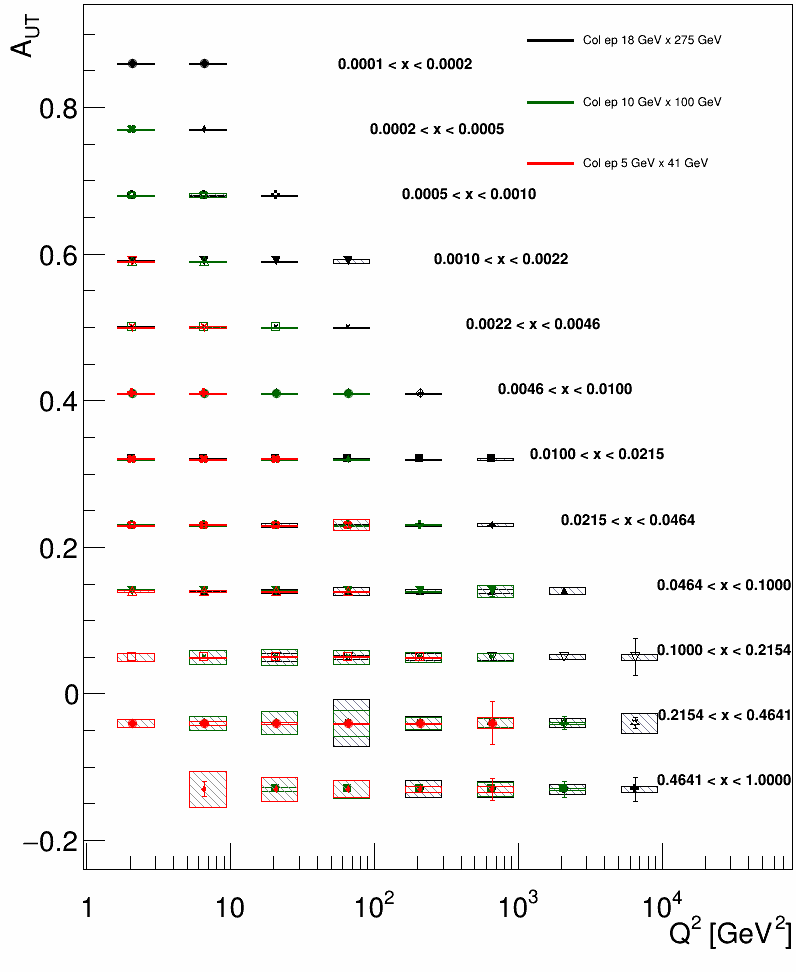}
    \caption{Projected Collins asymmetry uncertainties for all $x$ bins as a function of $Q^2$ for various collision energy combinations. For visibility, all $z$ and $P_T$ were combined for this figure. The asymmetries are offset for each $x$ bin while the uncertainty boxes represent the conservative estimates of the systematic uncertainties due to detector smearing as evaluated by the differences between true and reconstructed asymmetries.}
    \label{fig:heraevo}
\end{figure}

To illustrate a larger picture of the sensitivity that can be reached in terms of the TMD evolution, Fig.~\ref{fig:heraevo} highlights the whole $x$ and $Q^2$ range that can be covered for Sivers or Collins asymmetries, here displayed as a function of $Q^2$, where all $x$ bins are vertically offset. It becomes clear that while at high and low $x$ only the lowest/highest collision energies contribute, at the intermediate regions, the different energies overlap and extend the $Q^2$ lever arm. In this example the $z$ and $P_T$ dependence was integrated out for visibility reasons, but in the actual EIC data, the statistical precision is good enough to study this in a finer binning.

\section{Impact studies on Transversity, Sivers function and Collins function}
The uncertainties of the pseudo-data were then used in the two global fits described below to evaluate the change in the uncertainties when including the expected ECCE data. While the actual future data will require a completely new fit of all available data, for this impact study a re-weighting technique has been applied. The same central values of the parameterizations as in the prefious fits were used and only the change in resulting uncertainties is estimated. See, for example \cite{Gamberg:2021lgx} for a more detailed description.   

\subsection{Sivers function measurements:}
\begin{figure*}[ht]
    \centering
      \includegraphics[width=0.48\textwidth]{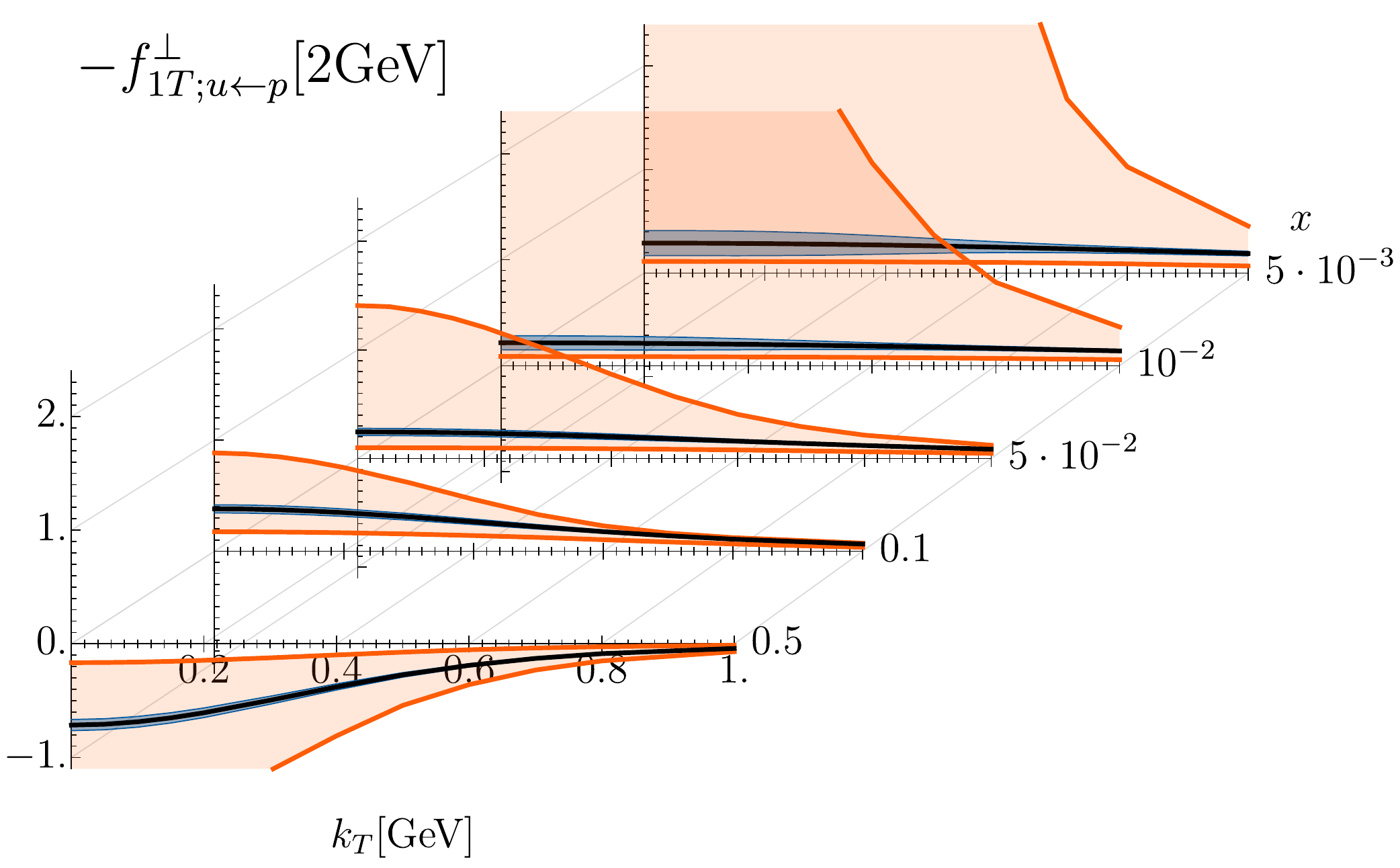}
   \includegraphics[width=0.48\textwidth]{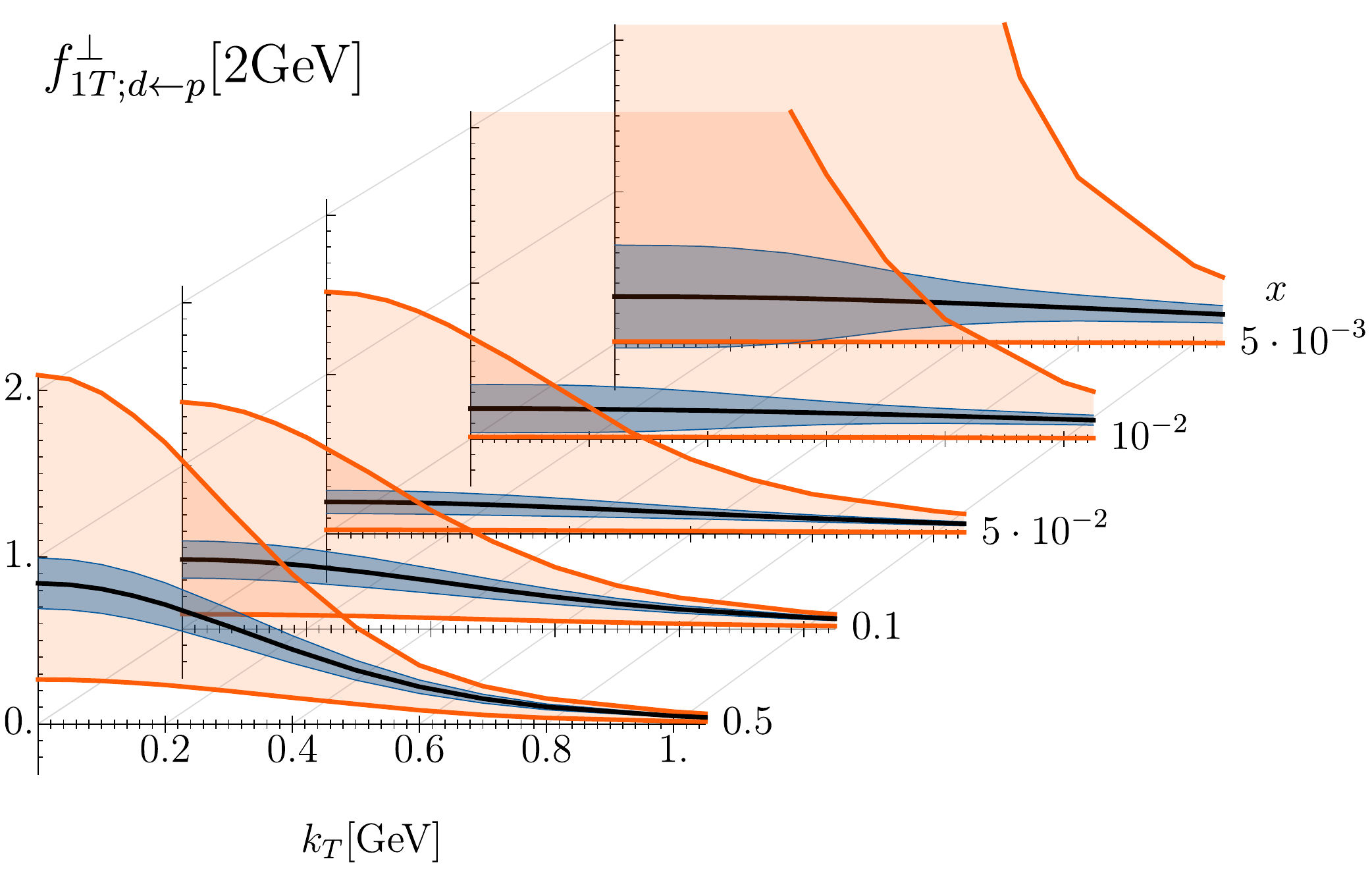}

    \caption{Expected impact on up (left) and down (right) quark Sivers distributions as a function of the transverse momentum $k_T$ for different values of $x$, obtained from SIDIS pion and kaon EIC pseudo-data, at the scale of 2 GeV. The orange-shaded areas represent the current uncertainty, while the blue-shaded areas are the uncertainties when including the ECCE pseudo-data.}
    \label{fig:part2-subS-PartStruct-MultiPart.udSivers}
\end{figure*}
\label{sec:part2-subS-SecImaging-TMD3d.transversity}

The impact studies based on these uncertainties can be seen in Fig.~\ref{fig:part2-subS-PartStruct-MultiPart.udSivers} for the up and down quark Sivers functions as a function of the intrinsic transverse momentum $k_T$ in various slices of $x$. They show the expected uncertainties of the up and down quark Sivers functions including the ECCE pseudo-data in comparison to the current knowledge as extracted from \cite{Bury:2021sue}. The central lines are fixed to those from the current data extraction which has slightly changed in comparison to the preliminary results that were the basis for the same figures in the Yellow Report \cite{EICYellowReport}. 

\begin{figure}[ht]
    \centering
   \includegraphics[width=0.48\textwidth]{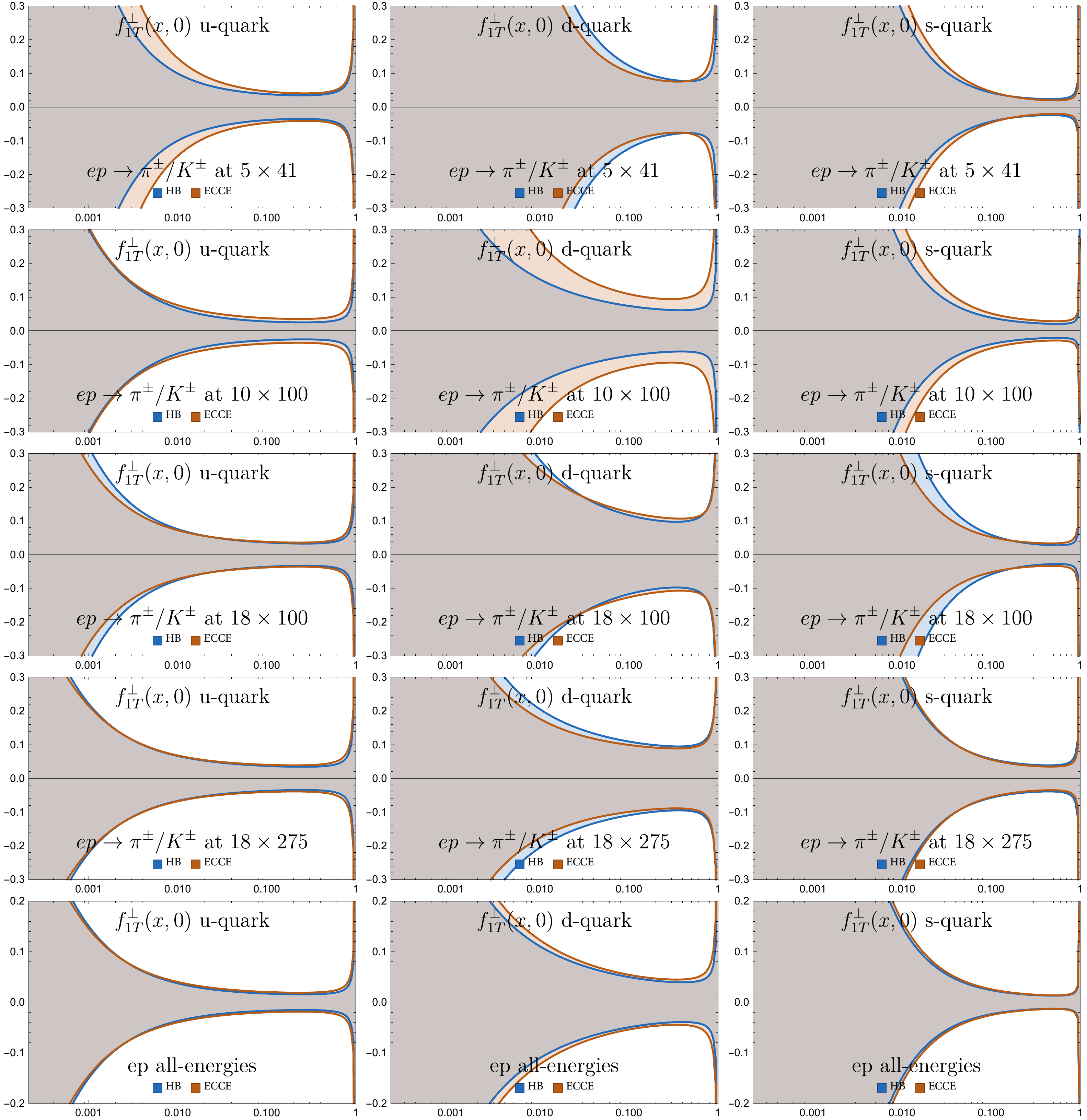}
    \caption{Expected impact on up (left), down (middle) and strange (right) quark Sivers distributions as a function of $x$ , obtained from SIDIS pion and kaon EIC pseudo-data, at the scale of 2 GeV. The blue-shaded areas represent the uncertainties obtained in the Yellow Report based on the parametrized reference detector, while the orange-shaded areas are the uncertainties when including the ECCE pseudo-data. From top to bottom the impact of the different collision energies is highlighted.}
    \label{fig:SiversHB_ECCE}
\end{figure}

The explicit comparison of the Sivers function uncertainties for up, down and strange quarks from ECCE pseudo-data and the parametrized reference detector of the Yellow Report are shown in Fig.~\ref{fig:SiversHB_ECCE} as a function of $x$. It shows that apart from slight differences due to the ranges assumed for the particle identification and the amount of actual detector smearing, the uncertainties are quite comparable. This again highlights that the ECCE detector concept fulfills the requirements set for the reference detector in the Yellow Report using realistic simulations of detectors, materials and support structure.

It should be noted that any impact figure relies on the baseline parameterization of that particular group of global fitters and the assumptions that group has used within their global fits. The expected uncertainties from other groups will look different, particularly in regions of low-$x$ where so far no single spin asymmetry data exists and most of the uncertainty bands originate in the functional form, positivity bounds and other constraints. The question is not so much about which of these fits is right as much as the fact that a similar increase in the precision with EIC or ECCE pseudo-data is expected. As such, the impact studies presented here are only one representative of the large number of groups working on global extractions for the Sivers functions \cite{Efremov:2004tp,Anselmino:2005ea,Collins:2005ie,Vogelsang:2005cs,Anselmino:2008sga,Bacchetta:2011gx,Sun:2013hua,Echevarria:2014xaa,Boglione:2018dqd,Luo:2020hki,Bacchetta:2020gko,Echevarria:2020hpy,Bury:2020vhj}.

\subsection{Collins-function-based transversity measurements:}

For the quark transversity distribution and the related tensor charges also the impact studies of the Yellow Report \cite{EICYellowReport,Gamberg:2021lgx} were revisited using the ECCE pseudo-data at the same energies as in the Yellow Report, following the methodology of the global fit from \cite{Cammarota:2020qcw}.

\begin{figure*}[ht]
    \centering
    \includegraphics[width=0.97\textwidth]{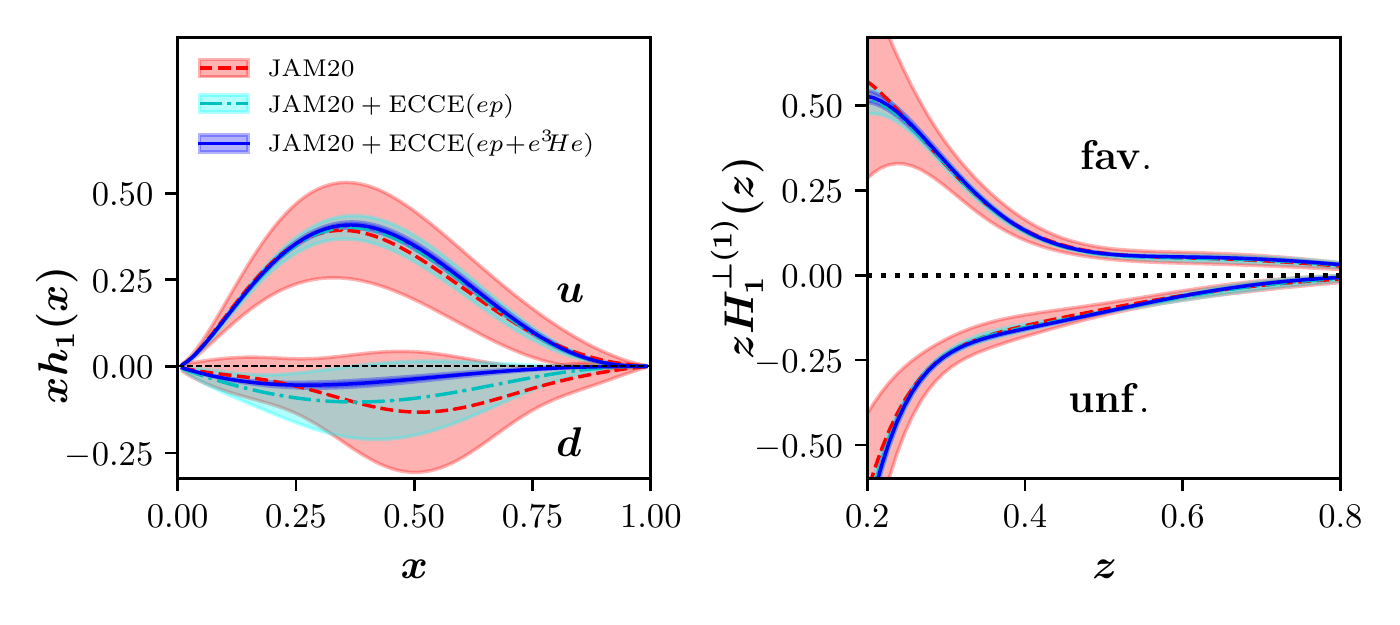}
    \includegraphics[trim={ 0 0 6cm 0},clip,width=0.8\textwidth]{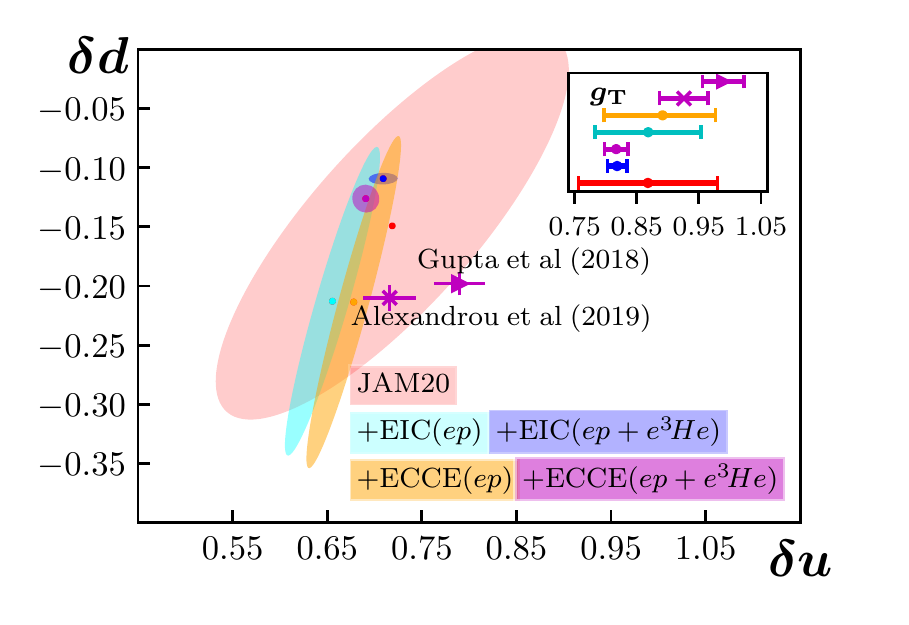}
    \caption{Top: Expected impact on the up and down quark transversity distributions and favored and un-favored Collins function first moment when including EIC Collins effect SIDIS pseudo-data from \mbox{\textit{e}+p} and \mbox{\textit{e}+He} collisions~\cite{Gamberg:2021lgx}. 
    Bottom: The impact on the up quark ($\delta u$), down quark ($\delta d$), and isovector ($g_T$) tensor charges from the ECCE pseudo-data and their comparison to the Yellow Report and lattice data \cite{Gupta:2018qil,Alexandrou:2019brg}.}
    \label{fig:fig:part2-subS-PartStruct-MultiPart.singlehadtransversity}
\end{figure*}

The resulting transversity distributions and tensor charges can be seen in Fig.~\ref{fig:fig:part2-subS-PartStruct-MultiPart.singlehadtransversity}. Again, one can see that the impact of the EIC data significantly reduced the uncertainties on the charges. Particularly the inclusion of the $^3$He data improves the so far poorly known down quark transversity distribution and tensor charge. The tensor charges, as well as their iso-vector combination $g_T$ can be precisely evaluated thanks to the large range in $x$ to reduce possible uncertainties due to the extrapolation to lower $x$ that currently is one of the main uncertainties in the extractions. One can also see that the impact of the ECCE pseudo-data is comparable to the impact studies that were performed using the parametrized reference detector of the Yellow Report. As such, it is again shown that ECCE fulfills the detector requirements to successfully obtain these physics goals. Again for this impact study, the expected impact is based on the assumptions, functional form, etc of this particular global fit, but the impact should be comparable for other transversity extractions as well. The small differences between Yellow Report and ECCE impact values arise from the different pseudo-data sets resulting is slightly different central values.
Most importantly, the expected uncertainties on these tensor charges are comparable with the lattice QCD simulations \cite{Gupta:2018qil,Alexandrou:2019brg}, and therefore will allow a quantitative comparison of experimental and lattice results, where one can explore the potential impacts in the case of discrepancies \cite{Courtoy:2015haa,Gao:2017ade}. 

Additionally, also the extraction of the tensor charge via di-hadron fragmentation functions as performed by \cite{Courtoy:2012ry,Radici:2015mwa,Radici:2016lam} is a viable alternative that can also serve as a cross check given the different sources of uncertainties, both experimentally and theoretically.

\section{Outlook of further studies}

These studies have shown that the ECCE detector is well suited to extract the single spin asymmetries needed to obtain a better knowledge of the Sivers function, transversity and the Collins function. The expected impact on these quantities is comparable with that estimated in the Yellow report and therefore fulfills the physics requirements of an EIC detector. These single spin asymmetries represent only the flagship measurements in terms of the transverse spin and momentum structure of the nucleon while many more single hadron asymmetries can shed light on various other spin and orbit correlations of partons in the nucleon. Also di-hadron fragmentation helps with several of these physics quantities, most notably transversity. Those have not been covered but are expected to show also improvements on a similar level discussed in the Yellow report. 

As the ECCE detector develops, the simulations will become even more realistic and the data analysis will also progress closer to that of actually taken experimental data. As such, some optimization in selection criteria and binnings will take place. Similarly, the increasingly more realistic detector responses for the particle identification and the smearing can be addressed in an actual unfolding which should improve the systematic uncertainties that are currently very crudely estimated at least in quality, and possibly also quantitatively.  As mentioned previously, also relying on calorimetry information to obtain the scattered lepton kinematics as well as using also hadronic DIS kinematic reconstruction methods in some areas of phase space will likely further improve the quality of the expected measurements. 
\section{Acknowledgements}
\label{acknowledgements}

We thank the EIC Silicon Consortium for cost estimate methodologies concerning silicon tracking systems, technical discussions, and comments.  We acknowledge the important prior work of projects eRD16, eRD18, and eRD25 concerning research and development of MAPS silicon tracking technologies.

We thank the EIC LGAD Consortium for technical discussions and acknowledge the prior work of project eRD112.

We thank (list of individuals who are not coauthors) for their useful discussions, advice, and comments.

We acknowledge support from the Office of Nuclear Physics in the Office of Science in the Department of Energy, the National Science Foundation, and the Los Alamos National Laboratory Laboratory Directed Research and Development (LDRD) 20200022DR.

This work was also partially supported by the National Science Foundation under grant No. PHY-2011763,
, Grant  No.~PHY-2012002 , the U.S. Department of Energy under contract No.~DE-AC05-06OR23177 under  which Jefferson  Science  Associates,  LLC,  manages and operates Jefferson Lab, and within the framework of the TMD Topical Collaboration.
\bibliographystyle{unsrturl}
\bibliography{refs}

\end{document}